\documentclass[%
 reprint,
 amsmath,amssymb,
 aps,
]{revtex4-2}

\usepackage{graphicx}
\usepackage{dcolumn}
\usepackage{bm}
\usepackage{placeins}

\begin{document}

\preprint{APS/123-QED}
\title{Polarization-Controlled Photon Mode Switching and Photon–Magnon Coupling in a Planar Cavity-Magnonic System}

\author{Abhishek Maurya$^{1}$}
\author{Sachin Verma$^{1}$}
\author{Bojong Kim$^{2}$}
\author{Biswanath Bhoi$^{1}$}
\email{Corresponding authors, E-mail: biswanath.phy@itbhu.ac.in}
\author{Rajeev Singh$^{1}$}
\email{rajeevs.phy@itbhu.ac.in}
\author{Sang-Koog Kim$^{2}$}
\email{sangkoog@snu.ac.kr}
\affiliation{$^{1}$Nano-Magnetism and Quantum Technology Lab,
 Department of Physics, Indian Institute of Technology (Banaras Hindu University) Varanasi, Varanasi - 221005, India.}

\affiliation{$^{2}$National Creative Research Initiative Center for Spin Dynamics and Spin-Wave Devices Nanospinics Laboratory, Research Institute of Advanced Materials Department of Materials Science and Engineering, Seoul National University, Seoul 08826, Republic of Korea}%

\begin{abstract}
This work presents polarization-selective photon--magnon coupling (PMC) in a planar cavity--magnonic platform consisting of an electric--LC resonator (ELCR) side-coupled to a microstrip transmission line and integrated with a yttrium iron garnet (YIG) thin film. The ELCR supports two orthogonal photon modes at $\sim 3.93~\text{GHz}$ and $\sim 5.73~\text{GHz}$, whose excitation efficiency and radiative damping are governed by the resonator orientation relative to the fixed microwave field polarization of the transmission line. Rotating the resonator enables controlled switching between these modes, thereby providing orientation-controlled tuning of photon--magnon hybridization. An equivalent circuit model is developed to describe the two orthogonal photon modes, incorporating their intrinsic and extrinsic damping arising from mutual inductive coupling between the microstrip line and the resonator, and successfully captures the polarization-driven mode switching. Furthermore, an effective three-mode Hamiltonian incorporating two photon modes of the ELCR and a magnon mode in the YIG film is formulated, accurately reproducing the observed photon--magnon coupled-mode evolution. The results reveal a pronounced angular tunability of the PMC strength, where the effective coupling is redistributed between two competing photon--magnon interaction channels. At $\theta = 0^\circ$, only the lower-frequency photon mode is excited, yielding a finite coupling of $g_{31} = 56.5~\text{MHz}$, while the higher-frequency mode remains inactive and does not participate in the hybridization. As the rotation angle increases, both channels become active, with the effective interaction redistributed between two photon--magnon channels over different angular ranges. In particular, $g_{31}$ increases from $56.5$ to $98~\text{MHz}$ over $0^\circ$--$60^\circ$ before vanishing at $90^\circ$, whereas $g_{23}$ emerges at finite angle and decreases from $76$ to $30~\text{MHz}$ over $30^\circ$--$90^\circ$. This evolution reflects a polarization-driven redistribution between two competing hybridization channels, yielding a measured transition near $25.7^\circ$ and a symmetry-related model-predicted transition near $154.3^\circ$. These findings establish resonator-orientation--driven polarization selectivity as a robust and versatile mechanism for controllable photon--magnon interactions in planar architectures, opening new avenues for waveguide-integrated magnonic devices and reconfigurable hybrid magnonic--photonic platforms.

\end{abstract}

\maketitle


\section{\label{sec:level1}INTRODUCTION}
The coherent interaction between microwave photons and collective spin excitations (magnons) in magnetic materials has emerged as a key building block for hybrid quantum technologies \cite{Xiang2013Hybrid,Zhang2015MagnonMemory,Wallquist2009Hybrid,Maurya2024PhotonMagnon}. Photon--magnon coupling (PMC) enables the transduction \cite{Bejarano2024MagnonTransduction}, storage, and processing of quantum information by combining the long coherence times \cite{Xu2023MagnomechanicalStorage} and tunability of magnetic excitations with the flexibility of photonic circuits \cite{MaierFlaig2017TunableCoupling}. Such hybrid systems provide promising platforms for quantum memories\cite{Heshami2016QuantumMemories}, coherent signal processing, and information transfer between disparate quantum systems \cite{Xu2021CoherentMagnonics}, particularly in the microwave domain relevant to superconducting and solid-state quantum technologies \cite{Burkard2020HybridQED}. Beyond quantum information processing, PMC also plays a central role in waveguide magnonics \cite{Qian2025ChiralMagnonics}, where controllable light--matter interactions enable on-chip signal routing \cite{Pechal2016Switch}, filtering, and nonreciprocal functionalities \cite{Lecocq2017JosephsonAmplifier,Wang2019Nonreciprocity}.

For practical quantum and magnonic applications, however, it is not sufficient to realize photon--magnon coupling alone; precise and reconfigurable control over the coupling pathways and participating modes is essential \cite{Rao2021PerfectAbsorption,Jeon2024DualCoupling,Khan2026DynamicControl}. In realistic photonic environments, resonators often support multiple photon modes \cite{Jeon2024DualCoupling}, each with distinct spatial field distributions and coupling characteristics. The ability to selectively address, enhance, or suppress specific photon--magnon interactions is therefore crucial for minimizing unwanted crosstalk, engineering mode-dependent functionalities, and scaling hybrid architectures \cite{Yang2025CoupledMagnons,Li2024PTCoupling}. Achieving such control in a simple, robust, and planar-compatible manner remains a central challenge in cavity magnonics.

Previous approaches to controlling PMC have primarily relied on tuning the resonance conditions of the constituent subsystems \cite{Bhoi2017PlanarCoupling,Bhoi2019AbnormalAnticrossing}. These include adjusting the external magnetic field to shift the magnon frequency, modifying cavity geometries to alter photon spectra \cite{Shrivastava2024PhotonPhoton}, or dynamically tuning resonators using varactors \cite{Girich2024NonlinearSRR}, superconducting elements \cite{Jouanny2025KineticInductance}, or mechanical actuation \cite{Ye2024StretchableSRR}. While effective, these methods often introduce additional complexity, dissipation, or limited tunability, and may require active control elements that are incompatible with scalable or low-loss architectures. Moreover, most existing schemes treat photon modes as fixed entities and focus on frequency matching, offering limited control over which photon mode couples to magnons when multiple cavity modes are present \cite{Zhang2021MultimodeMagnonics}.

An alternative and largely unexplored route to controlling photon--magnon interactions lies in exploiting the polarization degree of freedom of microwave fields \cite{Shi2014PolarizationRotator}. In planar transmission-line systems, the microwave electric and magnetic fields possess well-defined polarizations, and cavity photon modes can exhibit fundamentally different electric- and magnetic-dipole character depending on their symmetry and current distribution \cite{Hand2008ELCResonators}. Since magnons couple directly to the microwave magnetic field, while cavity excitation and radiative damping depend on both electric and magnetic field components, polarization provides a natural and powerful selection rule for controlling photon--magnon coupling. By tailoring how cavity modes project onto the fixed field polarizations of a waveguide, it becomes possible to selectively excite specific photon modes and, consequently, to control their coupling to magnons without altering the intrinsic properties of the cavity or the magnetic material \cite{Schurig2006NegativePermittivity}.

In this context, polarization-selective coupling based on resonator orientation offers a particularly simple and robust mechanism. Rather than dynamically tuning frequencies or modifying circuit elements, control is achieved purely through geometric alignment, which preserves the planar nature of the device and avoids additional loss channels. Importantly, this approach allows different photon modes of the same resonator—distinguished by their electric- or magnetic-dipole dominance—to be selectively addressed, switched, or suppressed, providing a new degree of freedom for engineering multimode photon--magnon systems.

In this work, we investigate polarization-selective photon--magnon coupling in a planar cavity system consisting of an electric--LC resonator (ELCR) side-coupled to a microstrip transmission line and integrated with a yttrium iron garnet (YIG) thin film. The ELCR supports two orthogonal photon modes arising from distinct current distributions in the resonator, each associated with different current-loop configurations. By studying ELCRs oriented at different angles with respect to the fixed microwave field polarization of the transmission line, we demonstrate controlled switching between these two modes through polarization-selective excitation. This selective mode excitation directly translates into tunable photon--magnon hybridization, enabling one coupling channel to be suppressed while the other is activated. To quantitatively describe the observed mode-switching behavior, an equivalent circuit model \cite{Pozar2021MicrowaveEngineering} is developed, along with an effective three-mode Hamiltonian formalism \cite{Harder2021CavityMagnonics} that captures the interaction of the two photon modes with magnons and reproduces the measured spectra well. Our results establish polarization-selective coupling as a useful approach for controlling photon--magnon interactions in planar cavity systems. This approach may enable further studies for waveguide-based magnonics, mode-selective hybrid devices, and reconfigurable quantum architectures where precise control over multimode interactions is essential.

\section{DESIGN AND PHOTON RESONANCES OF ELECTRIC–LC RESONATOR (ELCR)}
\subsection{Numerical Simulation of Polarization-Selective Photon Modes}
Figure 1(a) illustrates the planar photon resonator system investigated in this work, consisting of an electric--LC resonator (ELCR) placed adjacent to a microstrip transmission line in a planar geometry. The device geometry was first optimized through full-wave electromagnetic simulations and subsequently implemented experimentally.

Numerical simulations were carried out using CST Microwave Studio to analyze the photon resonances supported by the ELCR, the associated electromagnetic field distributions, and the polarization-dependent coupling between the resonator and the microstrip line. The simulated structure follows a three-layer planar configuration, as shown in Fig. 1(a). The bottom layer consists of a continuous copper ground plane with a thickness of 0.035 mm. Above it lies a dielectric substrate with a thickness of 0.64 mm, a relative permittivity of ($\varepsilon_r = 10.2$), and a dissipation factor of 0.0023 at 10 GHz. The top layer contains the patterned copper microstrip transmission line and the ELCR, both with metal thickness 0.035 mm. The microstrip width is chosen as 0.6 mm to realize a characteristic impedance of 50 Ohm. When driven at microwave frequencies, the microstrip supports a quasi-TEM mode that generates a localized and spatially nonuniform magnetic near field around the conductor together with an electric field terminating perpendicular to the ground plane. These near fields provide inductive and capacitive excitation pathways for the ELCR. The resonator is placed adjacent to the microstrip line with a lateral separation of $t = 0.5$ mm to enable efficient near-field coupling. The ELCR geometry is defined by an outer radius $d = 7$ mm, track width $w = 0.6$ mm, split gap $s = 0.4$ mm, metal thickness $b = 0.035$ mm, arm length $l = 1.5$ mm, and inner radius $r = 1$ mm.

\begin{figure*}
    \centering
    \includegraphics[width=\textwidth]{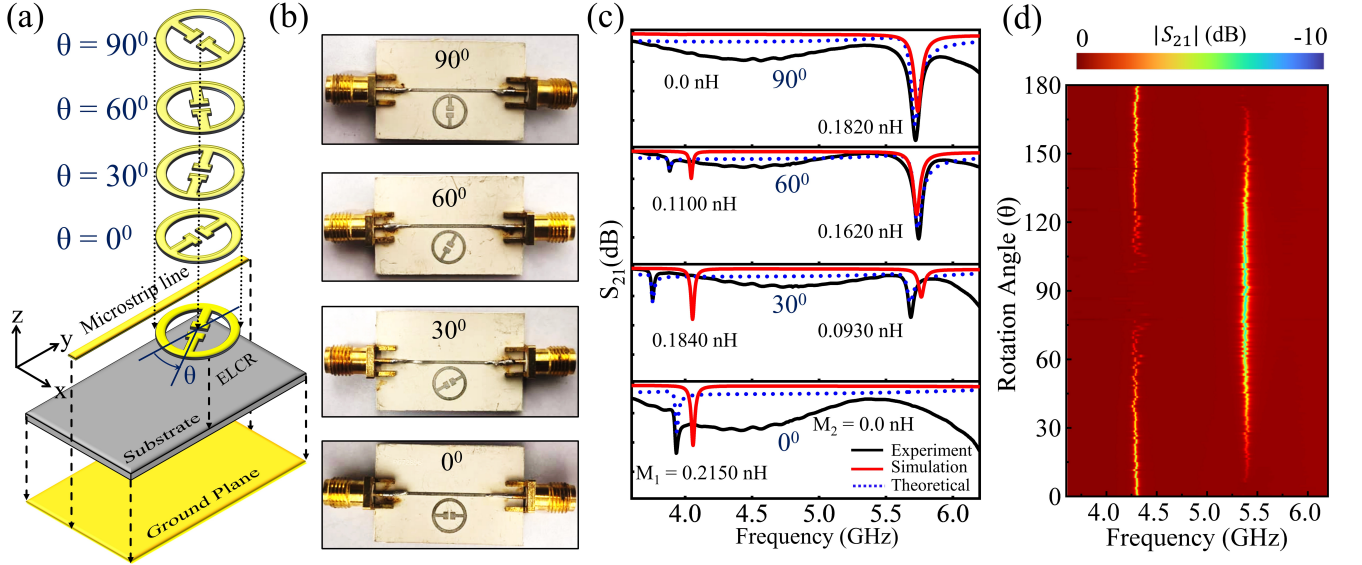}
    \caption{
(a) Schematic illustration of the simulation geometry used to investigate photon-mode excitation in the planar hybrid system. The electric--LC resonator (ELCR) is inductively coupled to a microstrip transmission line, whose ports (Port 1 and Port 2) are connected to a vector network analyzer (VNA) for transmission measurements. The structure is fabricated on a dielectric substrate with a continuous metallic ground plane on the backside. 
(b) Optical images of the fabricated ELCR--microstrip devices for $\theta = 0^\circ$, $30^\circ$, $60^\circ$, and $90^\circ$. 
(c) Transmission spectra $S_{21}$ for the four device orientations, showing experiment (black), CST Microwave Studio simulations (red), and equivalent-circuit fits (blue dotted curves). The values printed in each panel denote the extracted mutual inductances $M_1$ or $M_2$ associated with the active photon mode. 
(d) CST-simulated angle-dependent intensity map of the two photon resonances, illustrating complementary switching between photon mode-1 near 3.93~GHz and photon mode-2 near 5.73~GHz as the resonator is rotated.
}
    \label{F2}
\end{figure*}

To investigate polarization-dependent excitation, the ELCR was rotated by an angle $\theta$ with respect to the microstrip line, as illustrated in Fig. 1(a). The simulated transmission spectra $S_{21}$ for rotation angles $\theta = 0^\circ$, $30^\circ$, $60^\circ$, and $90^\circ$ are shown by the red curves in Fig. 1(c). Two well-separated photon resonances are observed: a lower-frequency mode near 3.93 GHz and a higher-frequency mode around 5.73 GHz. At $\theta = 0^\circ$, only the lower-frequency resonance is visible, which we identify as photon mode-1. As the resonator is rotated, the amplitude of this resonance progressively decreases, while a second resonance near 5.73 GHz gradually emerges. At $\theta = 90^\circ$, photon mode-1 becomes completely suppressed and photon mode-2 reaches its maximum intensity.
To further clarify the angular behavior, simulations were extended over the full rotation range $0^\circ \leq \theta \leq 180^\circ$. The resulting variation of the resonance intensities is summarized in Fig. 1(d). Photon mode-1 exhibits maximum intensity at $\theta = 0^\circ$, decreases monotonically with increasing angle, and vanishes near $\theta = 90^\circ$. Upon further rotation, it reappears and recovers its maximum intensity at $\theta = 180^\circ$. In contrast, photon mode-2 displays complementary behavior, with negligible intensity near $\theta = 0^\circ$, maximum intensity at $\theta = 90^\circ$, and suppression again near $\theta = 180^\circ$. This complementary angular dependence indicates polarization-selective excitation and rotation-controlled switching between two orthogonal photon modes. While these simulations clearly reveal the existence of polarization-controlled photon-mode switching, the underlying physical mechanism remains to be clarified \cite{Shi2014PolarizationRotator,Naqui2014ModeSuppression}.
\hspace{-0.5cm}
\subsection{Mechanism of Polarization-Dependent Mode Switching}
To gain deeper insight and elucidate the mechanism responsible for the rotation-dependent excitation of the photon modes, we calculated the surface current distributions of the ELCR at the resonance frequencies corresponding to photon mode-1 and photon mode-2 \cite{Maurya2024PhotonMagnon}.

We first consider photon mode-1, whose current distribution for representative rotation angles is shown in Fig. 2(a). The surface current pattern reveals two dominant circulating current loops flowing in opposite directions. These loops couple inductively to the spatially nonuniform microwave magnetic field generated by the microstrip line. Because the microwave magnetic near field generated by the microstrip line is spatially nonuniform, the two loops experience unequal magnetic flux and therefore contribute unequal inductive couplings of opposite sign \cite{Naqui2014ModeSuppression}. We denote these two contributions by $M_{1\theta}^{(1)}$ and $M_{1\theta}^{(2)}$. The effective mutual inductance associated with photon mode-1 can therefore be written as $M_1 = (M_{1\theta}^{(1)} + M_{1\theta}^{(2)})/2$ \cite{Naqui2015SensorsSRR}. At $\theta = 0^\circ$, the nonuniform magnetic field of the microstrip breaks the symmetry between the two loops such that $|M_{1\theta}^{(1)}| \neq |M_{1\theta}^{(2)}|$. As a result, a finite net mutual inductance $M_1$ arises, enabling efficient magnetic excitation of the resonator and producing the pronounced resonance near 3.93 GHz observed in the transmission spectrum. As the ELCR is rotated, the geometric symmetry of the resonator relative to the excitation field gradually increases, causing the two inductive contributions to approach equal magnitude. Consequently, the effective mutual inductance decreases continuously, leading to the progressive reduction of the resonance intensity. At $\theta = 90^\circ$, the two loops experience equal and opposite magnetic flux $M_{1\theta}^{(1)} = -M_{1\theta}^{(2)}$, resulting in complete cancellation of the inductive contributions and hence $M_1(90^\circ) = 0$. In this configuration, magnetic excitation of photon mode-1 is suppressed, and the mode becomes effectively dark, which is consistent with the absence of a resonance feature around 3.93 GHz in the transmission spectrum.

\begin{figure*}
    \centering
    \includegraphics[width=\textwidth]{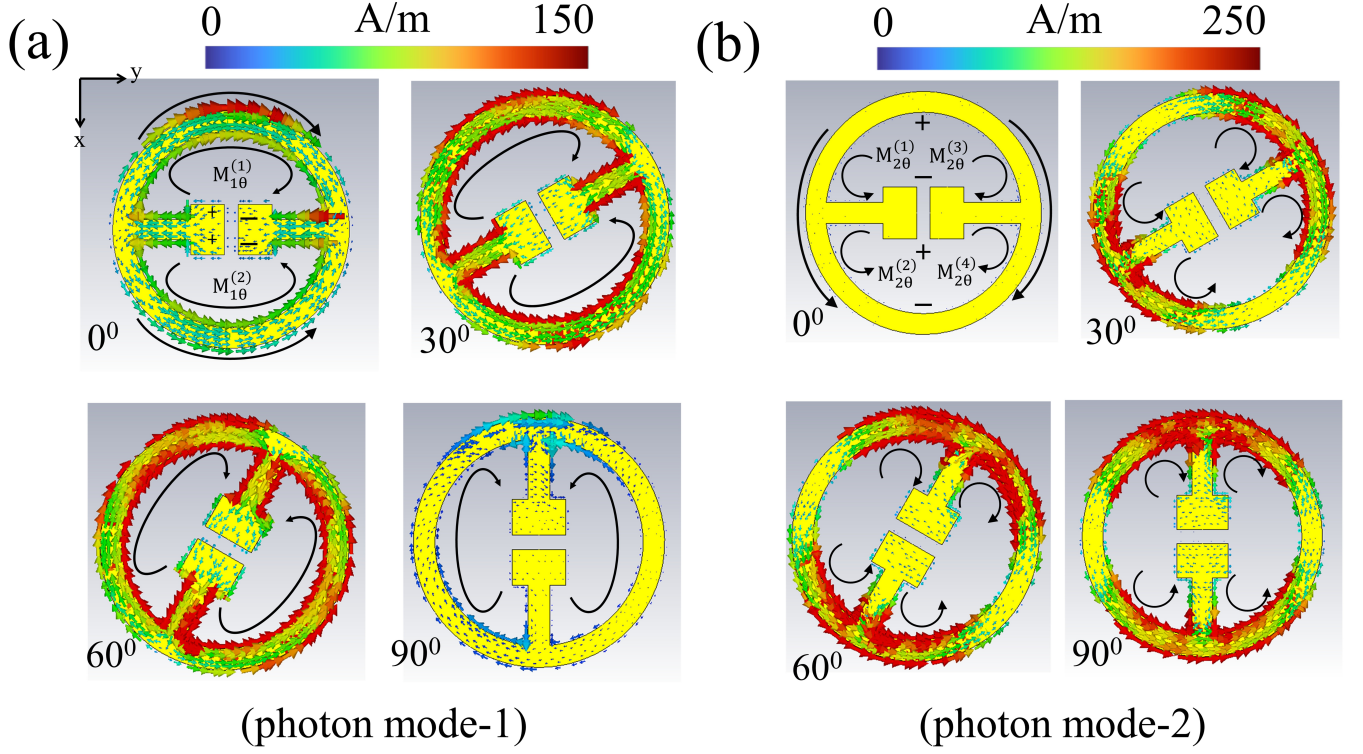}
    \caption{
Simulated surface-current distributions of the ELCR at the photon-mode resonances for different resonator orientations. 
(a) Photon mode-1, characterized by two counter-circulating current loops. The labels in the $0^\circ$ panel indicate the two loop contributions used to define the effective mutual inductance $M_1$. 
(b) Photon mode-2, characterized by four circulating current loops grouped into two pairs with opposite effective circulation. The labels in the $0^\circ$ panel indicate the loop contributions used to define the effective mutual inductance $M_2$. 
The current distributions are shown for $\theta = 0^\circ$, $30^\circ$, $60^\circ$, and $90^\circ$. Arrows indicate the current-circulation direction, and the color bars denote current magnitude.
}
    \label{F2}
\end{figure*}

The higher-frequency resonance at 5.73 GHz corresponds to photon mode-2. Although this mode is also excited via magnetic near-field coupling from the microstrip, its current topology is markedly different. As shown in Fig. 2(b), photon mode-2 is characterized by four circulating current loops distributed symmetrically around the resonator. Each loop interacts with the nonuniform microwave magnetic field and contributes a distinct inductive term. These loops can be grouped into two pairs with opposite effective circulation relative to the excitation field. We denote the contributions from one pair by $M_{2\theta}^{(1)}$ and $M_{2\theta}^{(2)}$, and those from the opposite pair by $M_{2\theta}^{(3)}$ and $M_{2\theta}^{(4)}$. The effective mutual inductance associated with photon mode-2 can therefore be expressed as 
$M_2 = \{(M_{2\theta}^{(1)} + M_{2\theta}^{(2)}) + (M_{2\theta}^{(3)} + M_{2\theta}^{(4)})\}/2$. 
At $\theta = 90^\circ$, the ELCR is oriented such that the symmetry between the two loop pairs is broken by the nonuniform magnetic field of the microstrip. As a result, the magnitudes of $(M_{2\theta}^{(1)} + M_{2\theta}^{(2)})$ and $(M_{2\theta}^{(3)} + M_{2\theta}^{(4)})$ are unequal, yielding a finite effective mutual inductance ($M_2(90^\circ) \neq 0$). This imbalance enables efficient magnetic excitation of photon mode-2, leading to the high-intensity resonance peak observed near 5.73 GHz in Fig. 2(b). For intermediate rotation angles ($0^\circ < \theta < 90^\circ$), partial symmetry is restored, and the inductive contributions from the four loops do not fully cancel. Consequently, the net mutual inductance remains finite but reduced, resulting in a gradual decrease in the resonance intensity, as confirmed by the surface current distributions and transmission spectra in Fig. 1(c). In contrast, at $\theta = 0^\circ$, the magnetic flux threading the four current loops becomes symmetric, such that $(M_{2\theta}^{(1)} + M_{2\theta}^{(2)}) = -(M_{2\theta}^{(3)} + M_{2\theta}^{(4)})$, leading to complete cancellation of the effective mutual inductance ($M_2(0^\circ) = 0$). In this configuration, magnetic coupling between the microstrip and the ELCR is suppressed, and photon mode-2 becomes dark, consistent with the absence of a resonance peak in the transmission spectrum at this orientation. These complementary coupling behaviors of photon mode-1 and photon mode-2 explain the polarization-controlled switching observed in the simulated transmission spectra.

\subsection{Equivalent-Circuit Model of the Two-Mode System}
To quantitatively describe the polarization-dependent excitation of the two photon modes, we develop an analytical model based on a lumped-element equivalent circuit representation of the system. This model captures the coupling between the microstrip transmission line and the two photon modes supported by the ELCR.

The equivalent circuit, shown in Fig. 3(a) and its simplified form in Fig. 3(b), consists of a microstrip transmission line coupled to two resonant modes. The transmission line is modeled by an inductance $L$ and a shunt capacitance $C$. The two photon modes are represented as independent RLC resonators. Photon mode-1 is described by inductance $L_1$, capacitance $C_1$, and resistance $R_1$, while photon mode-2 is characterized by $L_2$, $C_2$, and $R_2$. The resistances $R_1$ and $R_2$ account for intrinsic damping due to conductor and dielectric losses. Each resonator is coupled to the transmission line via mutual inductances $M_1$ and $M_2$, respectively. In addition, a mutual inductance $M_{12}$ is included to account for possible direct coupling between the two resonators.

\begin{figure*}
    \centering
    \includegraphics[width=\textwidth]{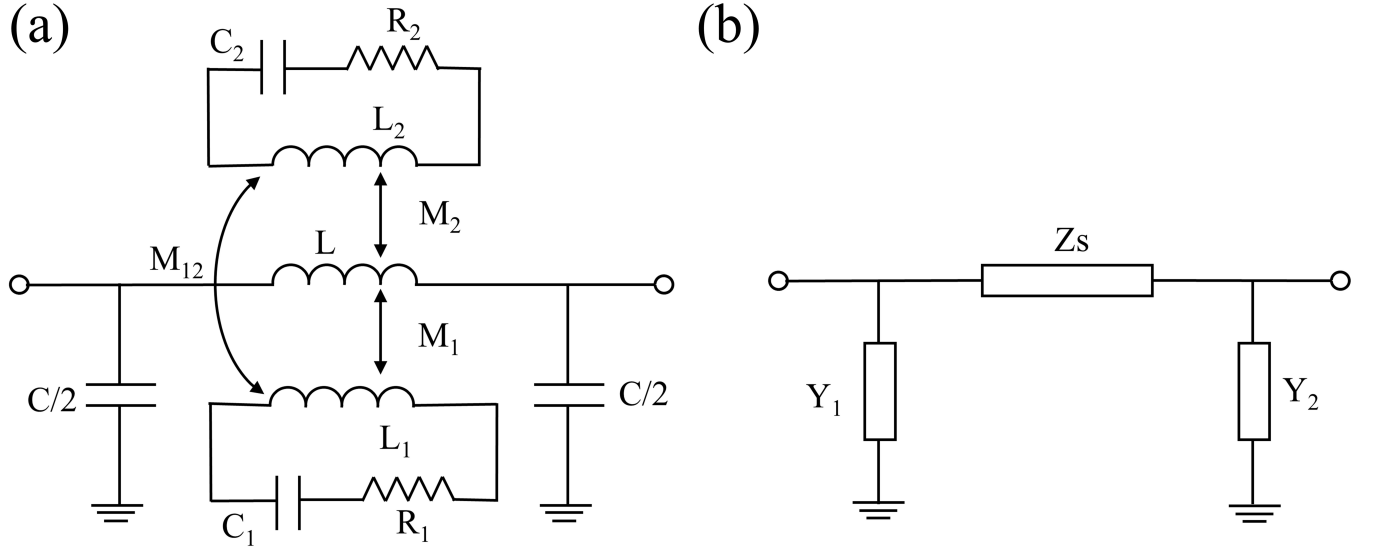}
    \caption{
Equivalent-circuit description of the photon-only ELCR--microstrip system. 
(a) Lumped-element circuit model of the microstrip line coupled to two ELCR photon modes represented by RLC resonators, with mutual inductive couplings $M_1$ and $M_2$ to the transmission line and a possible inter-mode mutual inductance $M_{12}$. 
(b) Reduced series--shunt equivalent circuit used in the ABCD-matrix analysis, where $Z_s$ is the effective series impedance of the coupled resonator system and $Y_1$ and $Y_2$ are the shunt admittances associated with the capacitive loading of the transmission line.
}
\label{F2}
\end{figure*}

Applying Kirchhoff’s circuit laws to this coupled system, the total series impedance seen by the transmission line can be written as (see supplementary material S2.3)
\begin{equation}
Z_s(\omega) = j\omega L + \Delta Z(\omega),
\tag{1}
\end{equation}
where $j\omega L$ represents the intrinsic inductive response of the transmission line, and $\Delta Z(\omega)$ captures the frequency-dependent contributions arising from the two coupled resonators and their mutual interactions, explicitly, the effective impedance of the coupled system, which is given as,
\begin{widetext}
\begin{align}
\Delta Z(\omega) &= 
\frac{
j\omega^3 \Big\{
C_1 M_1^2 \left(1 - \frac{\omega^2}{\omega_2^2}\right)
+ C_2 M_2^2 \left(1 - \frac{\omega^2}{\omega_1^2}\right)
+ 2\omega M_{12} M_1 M_2 C_1 C_2
+ j\omega C_1 C_2 (R_1 M_2^2 + R_2 M_1^2)
\Big\}
}{
\left(1 - \frac{\omega^2}{\omega_1^2} + j\omega R_1 C_1 \right)
\left(1 - \frac{\omega^2}{\omega_2^2} + j\omega R_2 C_2 \right)
- \omega^4 M_{12}^2 C_1 C_2
}
\tag{2}
\end{align}
\end{widetext}

Here, $\omega_1 = \frac{1}{\sqrt{L_1 C_1}}$ and $\omega_2 = \frac{1}{\sqrt{L_2 C_2}}$ denote the resonance frequencies of photon mode-1 and mode-2, respectively.

To compute the transmission coefficient $S_{21}$, we employ the standard ABCD-matrix formalism for cascaded two-port networks. The equivalent circuit can be decomposed into three elements: a central series impedance $Z_s$, and two identical shunt admittances $Y_1$ and $Y_2$, corresponding to the capacitive loading of the transmission line. These admittances are defined as $Y_1 = \frac{i\omega C}{2}$ and $ \qquad Y_2 = \frac{i\omega C}{2}.$ From the resulting ABCD parameters (see supplementary material S2.3), the transmission coefficient $S_{21}$ is calculated using standard relations for two-port networks \cite{Pozar2021MicrowaveEngineering,Kaur2016EITMagnonics,Aznar2008LeftHandedLines}:
\begin{equation}
S_{21} = \frac{2}{A + \frac{B}{Z_0} + C Z_0 + D}
\tag{3}
\end{equation}

Here,
\[
\begin{aligned}
A &= 1 + Y_2 Z_s, \quad
B = Z_s, \\
C &= Y_1 + Y_2 (1 + Y_1 Z_s), \quad
D = 1 + Y_1 Z_s.
\end{aligned}
\]

The above expression for $S_{21}$ applies to the two-resonator system, where $Z_0$ is the characteristic impedance of the transmission line; in the present case, $Z_0 = 50~\Omega$. The intrinsic damping ($\beta$) of each resonator is determined by its resistive losses and, under the near-resonance approximation, is given by photon mode-1 
$\beta_1 = \frac{\omega_1^2 R_1 C_1}{2}$
and photon mode-2 as 
$\beta_2 = \frac{\omega_2^2 R_2 C_2}{2}$
(see supplementary material S2.4) \cite{Naqui2015SensorsSRR}. 

Another important resonator parameter is the extrinsic damping, which quantifies radiative loss into the transmission line. Including the dependence on $Z_0$, the extrinsic damping is given by
$\gamma_1 = \frac{\omega_1^4 M_1^2 C_1}{2 Z_0}$ for photon mode-1 and $\gamma_2 = \frac{\omega_2^4 M_2^2 C_2}{2 Z_0}$ for photon mode-2 (see supplementary material S2.4) \cite{Naqui2015SensorsSRR}.

This analytical framework allows the measured transmission spectra to be directly related to the circuit parameters describing the resonator system. In particular, the mutual inductances $M_1$ and $M_2$ provide a quantitative measure of the radiative coupling strength of the resonator. In the following section, we use this model to analyze the experimentally measured spectra and extract the intrinsic and extrinsic damping parameters of the resonator.

\subsection{Experimental demonstration of Polarization-controlled Photon Mode Switching}

Based on the simulated design, ELCR devices with rotation angles of $0^\circ$, $30^\circ$, $60^\circ$, and $90^\circ$ relative to the microstrip line were fabricated using standard printed-circuit-board techniques. Transmission measurements were performed using a vector network analyzer (VNA) connected to the microstrip line, allowing direct measurement of the forward transmission coefficient $S_{21}$. The experimentally recorded $S_{21}$ transmission spectra are represented by black curves as shown in Fig 1(c). The experimental and simulation results are in good agreement except for a slight difference in the peak positions and the bandwidth of the resonant frequencies. These differences can be ascribed to the impedance mismatches resulting from the coaxial-to-microstrip line transitions at both connectors’ ends or mismatches in the dimensions of the samples with those of the model systems.

To quantitatively analyze the measurements, the experimental $S_{21}$ spectra were fitted using the analytical expression derived from the equivalent-circuit model [Eq. (3)]. The resulting fits are shown by the blue dotted curves in Fig. 1(c), demonstrating excellent agreement with the measured data. From these fits, the relevant circuit parameters for both photon modes were extracted and are summarized in Table 1 of the supplementary material. The extracted mutual inductances $M_1$ and $M_2$ reveal a pronounced angular dependence consistent with the simulation results. The $M_1$ associated with photon mode-1 decreases monotonically ($0.2150$ nH, $0.1840$ nH, $0.1100$ nH, and $0.00$ nH) with increasing rotation angle and vanishes at $\theta = 90^\circ$, whereas the $M_2$ of photon mode-2 increases ($0.00$ nH, $0.0930$ nH, $0.1620$ nH, and $0.1820$ nH) with rotation and reaches its maximum at $\theta = 90^\circ$. All other circuit parameters remain nearly constant, indicating that rotation modifies only the coupling between the excitation field and the resonator modes without altering the intrinsic properties of the ELCR. This opposite angular dependence of $M_1$ and $M_2$ arises from polarization-selective near-field coupling between the microstrip and the ELCR. Photon mode-1 corresponds to a current loop whose effective magnetic dipole is aligned with the microstrip magnetic field at $\theta = 0^\circ$, producing strong inductive coupling and a large $M_1$. As the resonator is rotated, the field overlap progressively decreases, causing $M_1$ to diminish and approach zero at $\theta = 90^\circ$, where the field becomes orthogonal to the current loop. In contrast, photon mode-2 involves four symmetric current loops whose coupling is minimal at $\theta = 0^\circ$ ($M_2 \approx 0$) because the magnetic field is nearly orthogonal to the dominant current paths. With increasing rotation angle, the field projection onto these loops increases, leading to a gradual enhancement of $M_2$ and maximum coupling near $\theta = 90^\circ$. This complementary behavior is consistent with the surface-current distributions discussed in the previous section.

This polarization-dependent coupling directly influences the energy dissipation pathways of the photon modes, particularly their radiative losses. The intrinsic damping rates of the two photon modes remain nearly constant ($\sim 11.0$ MHz and $\sim 25.0$ MHz for photon mode-1 and mode-2 respectively) across all rotation angles, reflecting losses determined primarily by the material properties and fixed resonator geometry. In contrast, the radiative damping scales with the square of the mutual inductance ($\gamma_{1,2} \propto M_{1,2}^2$) and exhibits a strong angular dependence, as confirmed in Fig. 4(a), where it increases monotonically with mutual inductance following an approximately quadratic trend. Notably, mode-2 ($\gamma_2$) shows a significantly steeper increase compared to mode-1 ($\gamma_1$), reflecting its higher radiative efficiency due to distinct current distributions.

\begin{figure*}
    \centering
    \includegraphics[width=\textwidth]{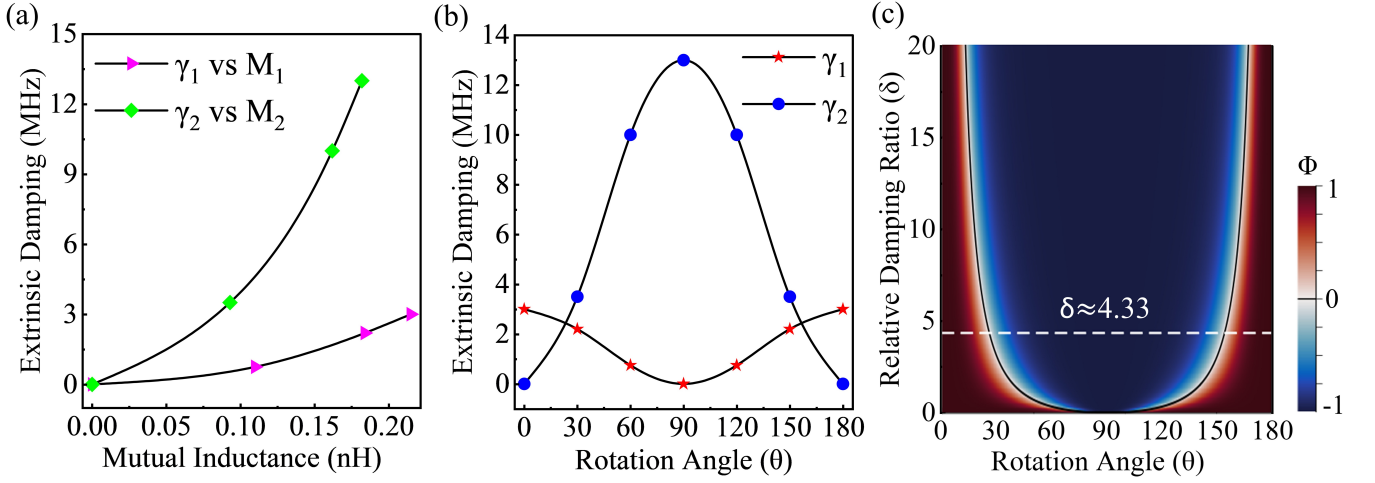}
    \caption{
Rotation-dependent radiative damping of the two photon modes. 
(a) Extracted extrinsic damping of photon mode-1 ($\gamma_1$) and photon mode-2 ($\gamma_2$) plotted as a function of their corresponding mutual inductances $M_1$ and $M_2$, showing an approximately quadratic dependence. 
(b) Angle dependence of the extrinsic damping rates of photon mode-1 ($\gamma_1$, red) and photon mode-2 ($\gamma_2$, blue), described by the projection relations $\gamma_1(\theta)=\gamma_1(0)\cos^2\theta$ and $\gamma_2(\theta)=\gamma_2(90^\circ)\sin^2\theta$. 
(c) Dissipation-transition map of the phenomenological order parameter $\Phi_\gamma$ as a function of rotation angle $\theta$ and relative damping ratio $\delta=\gamma_2(90^\circ)/\gamma_1(0)$. The black curve marks the boundary $\Phi_\gamma=0$, separating mode-1-dominated and mode-2-dominated radiative-loss regions. The white dashed line indicates the experimentally relevant damping ratio, $\delta\approx 4.33$.
}
\label{F2}
\end{figure*}

This angular dependence can be quantitatively understood from the projection of the excitation field onto the orthogonal current modes. Accordingly, the radiative damping follows
\begin{equation}
\gamma_1(\theta) = \gamma_1(0)\cos^2\theta, \qquad
\gamma_2(\theta) = \gamma_2(90^\circ)\sin^2\theta
\tag{4}
\end{equation}

Here, $\theta$ represents the effective polarization angle between the microstrip excitation field and the principal axes of the two photon modes. The parameters $\gamma_1(0) \approx 3$ MHz and $\gamma_2(90^\circ) \approx 13$ MHz correspond to the maximum radiative damping rates of mode-1 and mode-2, occurring at $\theta = 0^\circ$ and $\theta = 90^\circ$, respectively. The extrinsic damping, obtained by fitting the experimental data using an equivalent circuit model, exhibits a sinusoidal angular dependence, as shown in Fig. 4(b).

This behavior quantitatively explains the selective activation and suppression of the two orthogonal photon modes: alignment of a mode with the excitation field maximizes its radiative loss (bright state), whereas the orthogonal mode becomes weakly radiative (dark state) \cite{Zhang2015MagnonMemory}. Consequently, rotation of the ELCR enables polarization-controlled switching between the two excitation channels. To quantify this behavior, we introduce a dimensionless, phenomenological order parameter as a diagnostic quantity \cite{Singh2013Polarization,Born2019PrinciplesOptics}
\begin{equation}
\Phi_\gamma(\theta) = \frac{\gamma_1^2(\theta) - \gamma_2^2(\theta)}{\gamma_1^2(\theta) + \gamma_2^2(\theta)}
\tag{5}
\end{equation}

Substituting the angular dependence yields:
\begin{equation}
\Phi_\gamma(\theta) = \frac{\gamma_1^2(0)\cos^4\theta - \gamma_2^2(90^\circ)\sin^4\theta}{\gamma_1^2(0)\cos^4\theta + \gamma_2^2(90^\circ)\sin^4\theta}
\tag{6}
\end{equation}

To visualize this competition between dissipation channels, a two-dimensional transition map as shown in Fig. 4(c) is constructed as a function of $\theta$ and the relative damping ratio $\delta = \frac{\gamma_2(90^\circ)}{\gamma_1(0)},$
In terms of $\delta$, Eq. (6) can be rewritten as:
\begin{equation}
\Phi_\gamma(\theta) = \frac{\cos^4\theta - \delta^2 \sin^4\theta}{\cos^4\theta + \delta^2 \sin^4\theta}
\tag{7}
\end{equation}

To visualize the competition between the two dissipation channels, Fig. 4(c) is plotted using Eq. (7) as a function of the rotation angle $\theta$ and the relative damping ratio $\delta$. The color scale provides a visual representation of the dominant loss channel, with red (blue) regions indicating mode-1 (mode-2) dominance. The boundary defined by $\Phi_\gamma = 0$ separates these regimes and illustrates a continuous redistribution of radiative damping with rotation as the resonator polarization changes.

The experimentally determined damping ratio $\delta \approx 4.33$ defines a trajectory (white dashed horizontal line) along which the system evolves with $\theta$. Its intersection with the $\Phi_\gamma = 0$ boundary yields the critical switching angle $\theta_c \approx 25.7^\circ$, consistent with measurements. This transition (i.e. dissipation-channel switching) arises from polarization-dependent microwave excitation field projection onto orthogonal current loops, which modulates the radiation efficiency of each mode. As a result, the dominant leakage pathway can be continuously tuned between the two orthogonal resonator modes via simple geometric rotation. Importantly, the transition map demonstrates that this dissipation switching is robust over a wide range of $\delta$, indicating that it is primarily governed by geometric field--mode overlap. This polarization-controlled dissipation engineering provides a powerful and versatile mechanism for tailoring radiative losses and controlling microwave photon dynamics in planar cavity--magnonic systems, forming the basis for the photon--magnon interaction studies discussed in the following section.
\section{POLARIZATION-CONTROLLED PHOTON–MAGNON HYBRIDIZATION IN ELCR–YIG PLATFORM}
The polarization-selective excitation of the two orthogonal photon modes in the ELCR, established in the previous section through simulations and microwave transmission measurements, provides a controllable photonic platform for engineering light–matter interactions. To exploit this capability, we integrate the ELCR with a ferrimagnetic yttrium iron garnet (YIG) film, whose collective spin excitation (magnon mode) can be tuned via an external magnetic field. This hybrid configuration enables coherent photon–magnon coupling mediated by the two orthogonal photon modes of the resonator.

\subsection{. Experimental Realization of the ELCR–YIG Hybrid Platform}
To realize the hybrid system, a thin YIG film is placed directly above the ELCR resonator, as illustrated in Fig. 5(a). In this configuration, the YIG film interacts with the resonantly enhanced microwave magnetic field generated by the circulating current loops of the ELCR.

An in-plane static magnetic field $H$ is applied using an electromagnet. The ferromagnetic-resonance frequency of the YIG film follows the Kittel relation $f_m = \left(\frac{\gamma}{2\pi}\right)\sqrt{H(H + 4\pi M_s)},$
where $\gamma/2\pi = 2.8~\text{MHz/Oe}$ and $4\pi M_s \approx 1750~\text{G}$ for YIG \cite{Goldman2006Ferrites,Satapathy2023YIGFilms}. By varying the external field, the magnon mode can be tuned into resonance with either of the two photon modes. This tunability provides a controllable platform to investigate polarization-dependent photon--magnon interactions. With this experimental configuration established, we next examine how the photon--magnon interaction manifests in the microwave transmission spectra as the magnon mode is tuned through the photon resonances.

\begin{figure*}
    \centering
    \includegraphics[width=\textwidth]{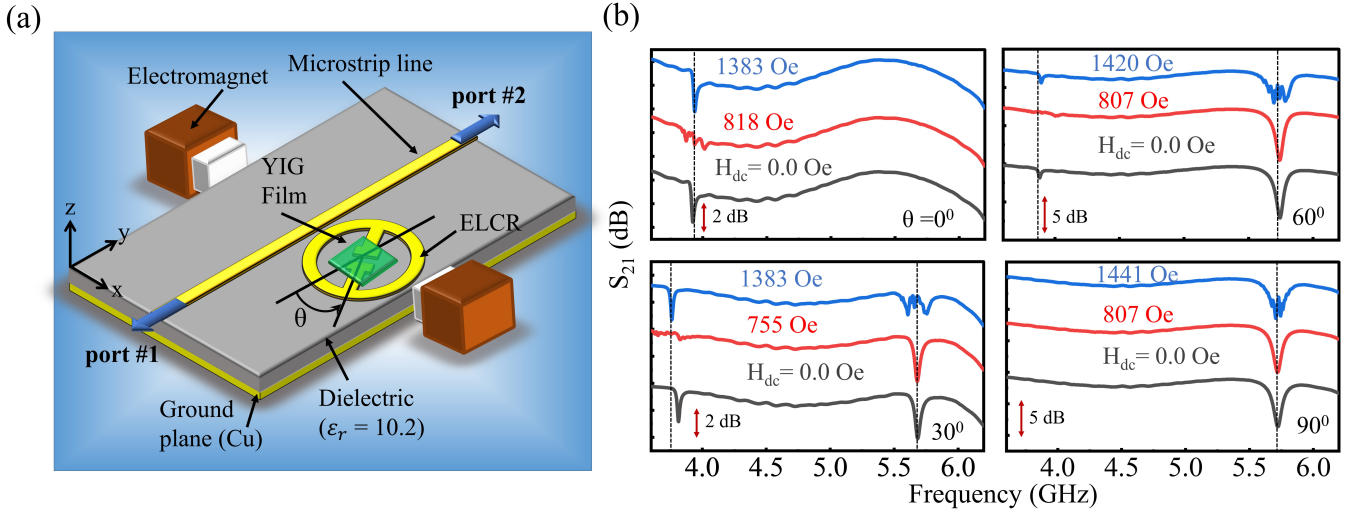}
    \caption{
Polarization-controlled photon--magnon coupling in the ELCR--YIG platform. 
(a) Schematic of the experimental configuration. A YIG film is placed above the ELCR, and an in-plane static magnetic field $H$ is applied to tune the magnon mode through the photon resonances. 
(b) Representative measured transmission spectra $S_{21}$ for $\theta = 0^\circ$, $30^\circ$, $60^\circ$, and $90^\circ$. In each panel, the black curve corresponds to the photon-only response at $H = 0$, while the colored curves correspond to selected magnetic fields at which the magnon mode is tuned close to photon mode-1 or photon mode-2. The traces are vertically offset for clarity and show angle-dependent hybridization between the magnon and the photon modes.
}
\label{F2}
\end{figure*}
    
\subsection{Angle-Dependent Anticrossing between Photon and Magnon Modes}
Figure 5(b) shows the experimentally measured transmission spectra $|S_{21}|$ as a function of frequency and magnetic field $H$ for different rotation angles of the ELCR relative to the microstrip excitation. As the magnetic field is varied, the magnon mode sweeps across the photon-mode frequencies of the resonator. For $\theta = 0^\circ$, the brown curve corresponds to the photon-only response measured at $H = 0~\text{Oe}$, where photon mode-1 appears at 3.93 GHz. Upon applying a magnetic field of $H = 818~\text{Oe}$, the magnon mode becomes resonant with photon mode-1, leading to the appearance of two well-resolved dips in the transmission spectrum, as shown by the red curve. When the magnon resonance is tuned far away from photon mode-1, the transmission spectrum (blue curves) recovers a single dip. Since photon mode-2 is absent at $\theta = 0^\circ$, no coupling involving this mode is observed. 

On the other hand, for $\theta = 30^\circ$, the photon-only spectrum at $H = 0~\text{Oe}$ (brown curve) exhibits two resonances corresponding to photon mode-1 at 3.76 GHz and photon mode-2 at 5.68 GHz. When the magnon mode is tuned to photon mode-1 at $H = 755~\text{Oe}$, two dips appear in the transmission spectrum, although with reduced intensity compared to the $\theta = 0^\circ$ case. In contrast, when the magnon frequency is tuned to photon mode-2 at $H = 1383~\text{Oe}$, a more pronounced splitting is observed, indicating stronger coupling to photon mode-2. A similar behavior is observed for $\theta = 60^\circ$, where photon mode-1 and photon mode-2 appear at 3.88 GHz and 5.73 GHz, respectively. Coupling between the magnon mode and photon mode-1 occurs at $H = 807~\text{Oe}$, while coupling to photon mode-2 is observed at $H = 1420~\text{Oe}$, as indicated by the red and blue curves in Fig. 5(b). However, at $\theta = 90^\circ$, photon mode-1 is completely suppressed, and only photon mode-2 remains visible at 5.73 GHz. In this configuration, coupling between the magnon mode and photon mode-2 is observed at $H = 1441~\text{Oe}$. When the magnon frequency approaches either photon mode, a clear mode splitting behaviour emerges in the transmission spectrum, indicating coherent photon--magnon hybridization. The slight variation in photon resonance frequencies with rotation angle is attributed to minor fabrication-induced asymmetries and alignment deviations during ELCR rotation \cite{Maurya2024PhotonMagnon}.

In order to examine the coupling behavior shown in Fig. 5(b), the $|S_{21}|$ transmission power in the $f$--$H$ plane was replotted for different rotation angles $\theta$, as shown in Fig. 6(a). For $\theta = 0^\circ$, only photon mode-1 is observed, and a clear anticrossing feature appears when it interacts with the magnon mode. As the rotation angle increases to $\theta = 30^\circ$, the intensity of photon mode-1 decreases, while photon mode-2 emerges and also exhibits an anticrossing feature with the magnon mode. At $\theta = 60^\circ$, the intensity of photon mode-1 is further reduced, whereas photon mode-2 becomes more prominent, with both modes continuing to display hybridization features. Finally, at $\theta = 90^\circ$, photon mode-1 is fully suppressed, while photon mode-2 reaches its maximum intensity and retains a visible anticrossing with the magnon mode.

\begin{figure}
    \centering
    \includegraphics[width=\linewidth]{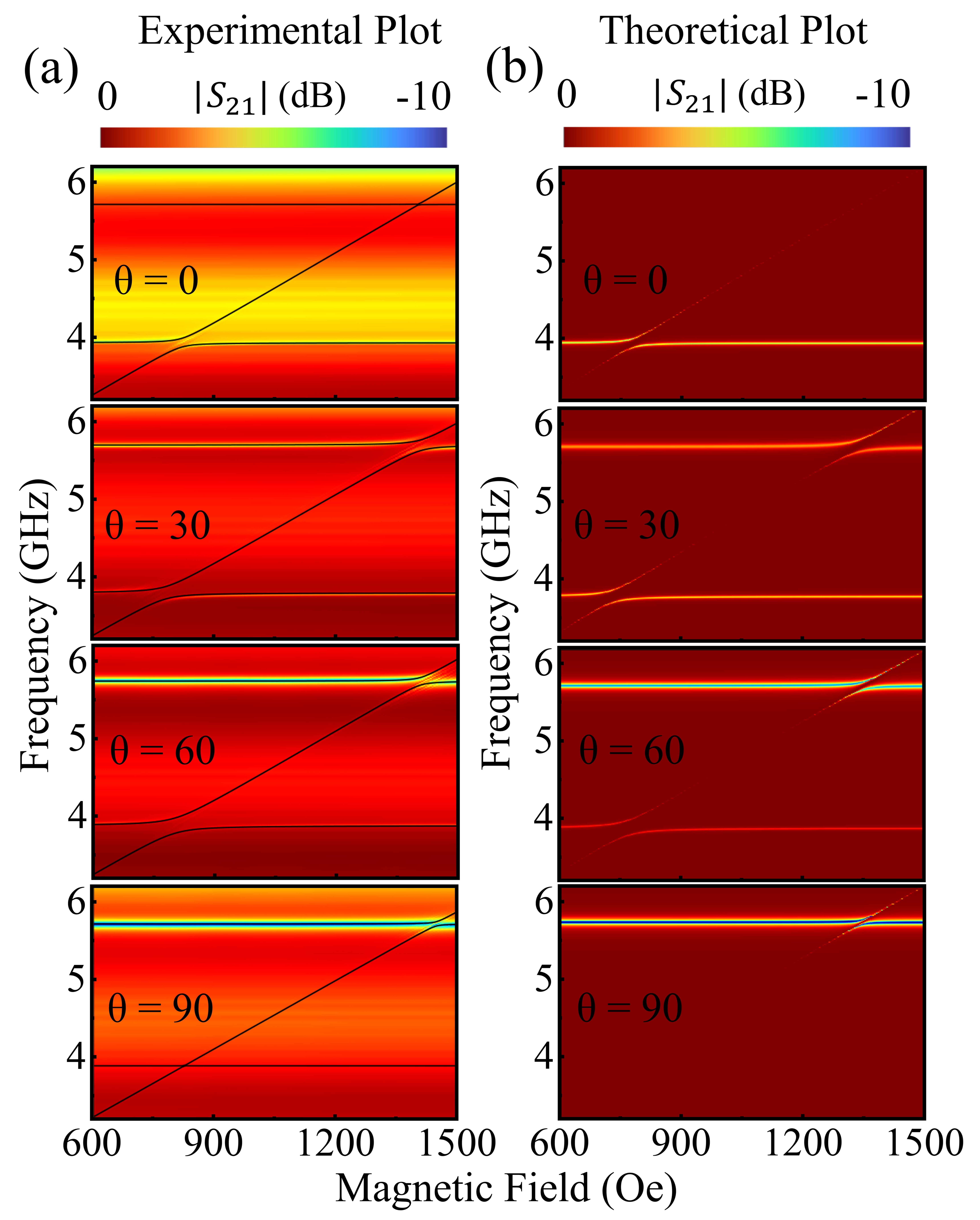}
    \caption{
Experimental and calculated photon--magnon hybridization in the frequency--field plane. 
(a) Measured transmission power $|S_{21}|$ in the frequency--magnetic-field ($f$--$H$) plane for the ELCR--YIG hybrid system at $\theta = 0^\circ$, $30^\circ$, $60^\circ$, and $90^\circ$. The overlaid black guide curves indicate guide trajectories extracted from the fitted eigenvalues of the effective three-mode coupling matrix [Eq.~(11)]. 
(b) Calculated transmission power $|S_{21}|$ obtained from Eq.~(13) for the same four rotation angles, showing good agreement with the measured angle-dependent redistribution of photon--magnon hybridization between the two photon channels.
}
\label{F2}
\end{figure}

Overall, the appearance and prominence of the anticrossing features depend strongly on the rotation angle of the ELCR. Rotation modifies the polarization overlap between the excitation field and the photon modes, leading to a redistribution of the hybridization signatures between the two photon--magnon interaction channels. These observations demonstrate that the photon--magnon interaction in the ELCR--YIG system can be controlled through polarization engineering. To quantitatively understand the underlying mechanism and extract the relevant coupling parameters, we develop a theoretical model describing the interaction between the two photon modes and the magnon mode.

\subsection{Effective Three-Mode Hamiltonian Model for Polarization-Selective Photon–Magnon Interaction}

To describe the observed hybridization behavior, the ELCR--YIG system is modeled as a three-mode cavity magnonic platform consisting of two orthogonal photon modes interacting with a single magnon mode. This configuration captures the essential physics of polarization-selective photon--magnon interactions governed by the resonator geometry. As illustrated in Fig. 7(a), a microwave photon propagating along the microstrip transmission line couples to three resonant modes with distinct frequencies. Two of these modes originate from the ELCR (shown as red spheres) and are therefore referred to as photon modes, whose resonance frequencies remain independent of the applied magnetic field. The third mode arises from the YIG film (shown as a green sphere) and exhibits a strong magnetic-field dependence, identifying it as the magnon mode \cite{Bhoi2017PlanarCoupling}. The dynamics of this hybrid system are described using a quantum input--output formalism that accounts for the intrinsic and radiative losses of the photon modes as well as the damping of the magnon mode. This framework provides an analytical description of the transmission response of the system and enables direct comparison with the experimentally measured spectra.

\begin{figure}
    \centering
    \includegraphics[width=\linewidth]{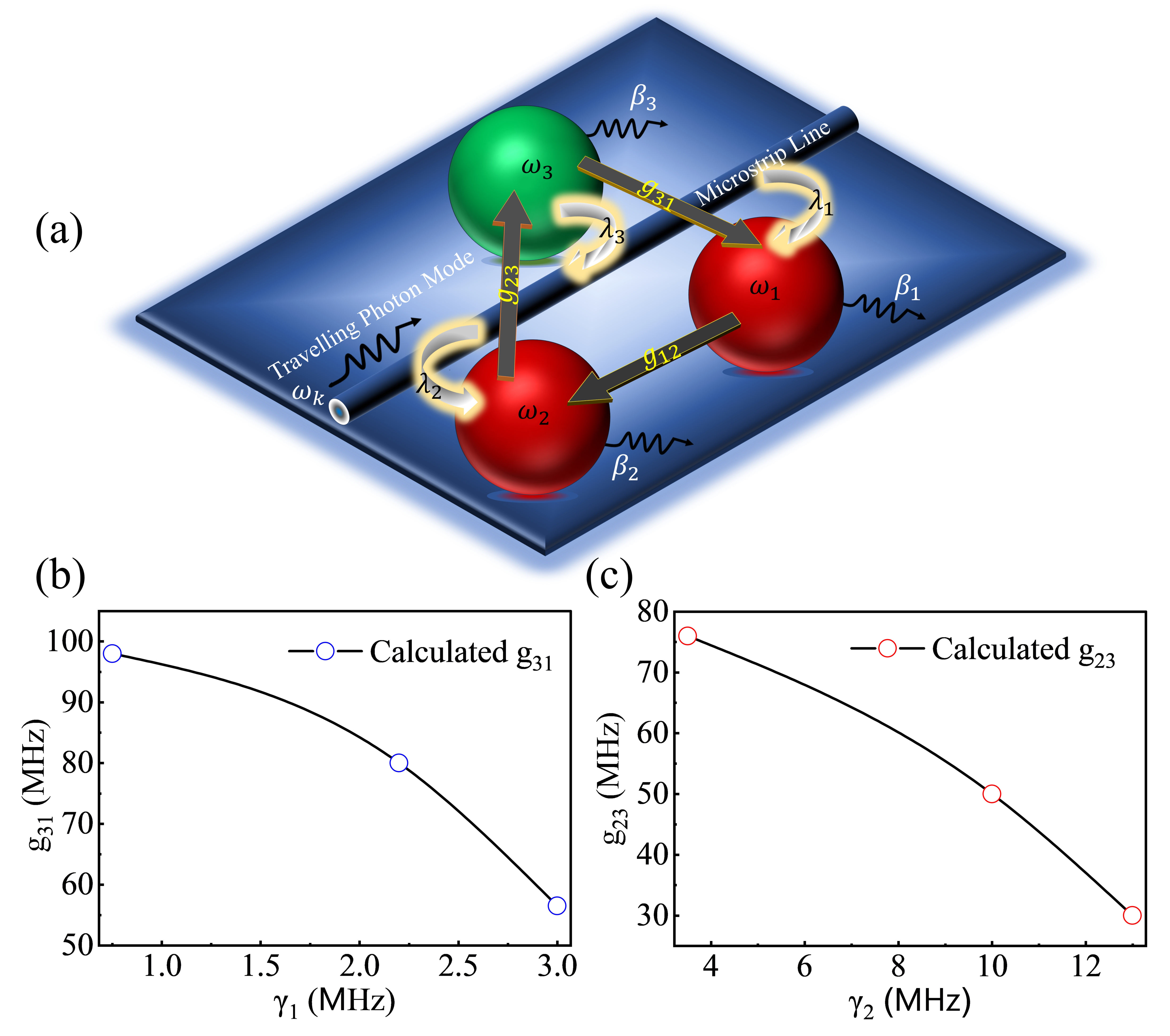}
    \caption{
Effective three-mode description of the polarization-controlled hybrid system and extracted coupling--damping trends. 
(a) Schematic of the three-mode model consisting of two ELCR photon modes and one YIG magnon mode coupled to a common microstrip continuum. Here, $g_{ij}$ denote coherent intermode coupling strengths, $\beta_i$ the intrinsic damping rates, and $\lambda_i$ the couplings to the waveguide continuum. 
(b) Extracted photon--magnon coupling strength $g_{31}$ associated with photon mode-1 plotted as a function of the extrinsic damping rate $\gamma_1$. 
(c) Extracted photon--magnon coupling strength $g_{23}$ associated with photon mode-2 plotted as a function of the extrinsic damping rate $\gamma_2$.
}
\label{F2}
\end{figure}

The three-mode coupled system can be described by the Hamiltonian \cite{Zhang2016Magnomechanics,Harder2021CavityMagnonics}:
\begin{widetext}
\begin{align}
\frac{H}{\hbar} &= \tilde{\omega}_1 a^\dagger a + \tilde{\omega}_2 b^\dagger b + \tilde{\omega}_3 c^\dagger c \nonumber + g_{12} (a + a^\dagger)(b + b^\dagger) + g_{23} (b + b^\dagger)(c + c^\dagger)
+ g_{31} (c + c^\dagger)(a + a^\dagger) \nonumber
+ \\&\int_{-\infty}^{+\infty} \omega_k p_k^\dagger p_k \, dk \nonumber + \int_{-\infty}^{+\infty}
\Big[
\lambda_1 \cos\theta (a + a^\dagger)(p_k + p_k^\dagger)
+ \lambda_2 \sin\theta (b + b^\dagger)(p_k + p_k^\dagger)
+ \lambda_3 (c + c^\dagger)(p_k + p_k^\dagger)
\Big] dk
\tag{8}
\end{align}
\end{widetext}

where $a$ ($a^\dagger$), $b$ ($b^\dagger$), and $c$ ($c^\dagger$) represent the annihilation (creation) operators associated with the photon mode-1, photon mode-2, and magnon mode, respectively. The resonance frequencies of the uncoupled modes are represented by the complex quantities $\tilde{\omega}_1$, $\tilde{\omega}_2$, and $\tilde{\omega}_3$, defined as $\tilde{\omega}_1 = \omega_1 - i\beta_1$, $\tilde{\omega}_2 = \omega_2 - i\beta_2$, and $\tilde{\omega}_3 = \omega_3 - i\beta_3$. Here, $\beta_1$, $\beta_2$, and $\beta_3$ represent the intrinsic damping rates of photon mode-1, photon mode-2, and the magnon mode, respectively. Since the three modes coexist on the same planar platform, their mutual interactions must be included. Accordingly, the next terms in the Hamiltonian describe the coupling between the modes with strengths $g_{12}$, $g_{23}$, and $g_{31}$, which are treated as effective coupling parameters within the input--output framework. The following term accounts for the continuum of traveling photons in the microstrip transmission line, represented by the bosonic operators $p_k$ ($p_k^\dagger$), which satisfy the commutation relation $[p_k, p_{k'}^\dagger] = \delta(k - k')$, where $\omega_k$ is the frequency corresponding to wave vector $k$. These traveling modes interact with the localized photon and magnon modes of the system. The final term describes the coupling of the localized modes to the waveguide photons with strengths $\lambda_1$, $\lambda_2$, and $\lambda_3$. All three modes are driven by a common traveling input field $p_{\mathrm{in}}$. While the magnon coupling $\lambda_3$ is assumed constant, the photonic couplings depend on the polarization overlap between the excitation field and the ELCR modes. Owing to the planar geometry, this overlap is controlled by the rotation angle $\theta$, resulting in effective coupling strengths $\lambda_1 \cos\theta$ and $\lambda_2 \sin\theta$ for the two photon modes.

Since the above Hamiltonian contains non-conserving terms, we apply the rotating-wave approximation (RWA) \cite{Zhang2016Magnomechanics}, under which the Hamiltonian reduces to
\begin{widetext}
\begin{align}
\frac{H}{\hbar} &= \tilde{\omega}_1 a^\dagger a + \tilde{\omega}_2 b^\dagger b + \tilde{\omega}_3 c^\dagger c \nonumber + g_{12} (a^\dagger b + a b^\dagger)
+ g_{23} (b^\dagger c + b c^\dagger)
+ g_{31} (c^\dagger a + c a^\dagger) \nonumber + \int_{-\infty}^{+\infty} \omega_k p_k^\dagger p_k \, dk \nonumber \\&+ \int_{-\infty}^{+\infty}
\Big[
\lambda_1 \cos\theta (a^\dagger p_k + a p_k^\dagger)
+ \lambda_2 \sin\theta (b^\dagger p_k + b p_k^\dagger)
+ \lambda_3 (c^\dagger p_k + c p_k^\dagger)
\Big] dk
\tag{9}
\end{align}
\end{widetext}

Transmission measurements are performed by connecting the microstrip to a vector network analyzer (VNA). The device is probed by the input fields $p_{\mathrm{in}}$ and the output fields $p_{\mathrm{out}}$ at the frequency $\omega$. Using coupled-mode theory, the equations of motion are given by (see supplementary material S3.3)
\begin{equation}
\frac{d}{dt}
\begin{pmatrix}
a(t) \\ b(t) \\ c(t)
\end{pmatrix}
=
-i H_{\mathrm{coupling}}
\begin{pmatrix}
a(t) \\ b(t) \\ c(t)
\end{pmatrix}
-i
\begin{pmatrix}
\sqrt{\gamma_1} \\ \sqrt{\gamma_2} \\ \sqrt{\gamma_3}
\end{pmatrix}
p_{\mathrm{in}}(t)
\tag{10}
\end{equation}

where
\begin{equation}
H_{\mathrm{coupling}} =
\begin{pmatrix}
\tilde{\omega}_1' & g_{12} - i\sqrt{\gamma_1 \gamma_2} & g_{31} - i\sqrt{\gamma_3 \gamma_1} \\
g_{12} - i\sqrt{\gamma_1 \gamma_2} & \tilde{\omega}_2' & g_{23} - i\sqrt{\gamma_2 \gamma_3} \\
g_{31} - i\sqrt{\gamma_3 \gamma_1} & g_{23} - i\sqrt{\gamma_2 \gamma_3} & \tilde{\omega}_3'
\end{pmatrix}
\tag{11}
\end{equation}

Here $\tilde{\omega}_1'$, $\tilde{\omega}_2'$, and $\tilde{\omega}_3'$ are the resultant complex frequencies, which include the extrinsic damping of the three modes, respectively, i.e. 
$\tilde{\omega}_1' = \omega_1 - i(\beta_1 + \gamma_1)$, 
$\tilde{\omega}_2' = \omega_2 - i(\beta_2 + \gamma_2)$, and 
$\tilde{\omega}_3' = \omega_3 - i(\beta_3 + \gamma_3)$. 
The corresponding extrinsic damping rates are 
$\gamma_1 = 2\pi \lambda_1^2 \cos^2\theta$, 
$\gamma_2 = 2\pi \lambda_2^2 \sin^2\theta$, and 
$\gamma_3 = 2\pi \lambda_3^2$ 
for photon mode-1, photon mode-2, and the magnon mode, respectively. 

Extrinsic damping can be written in the form of a rotation angle ($\theta$) dependence as 
$\gamma_1 = \gamma_1(0)\cos^2\theta$ and 
$\gamma_2 = \gamma_2(90^\circ)\sin^2\theta$ 
for the photon mode-1 and photon mode-2, respectively. 

The coupling matrix ($H_{\text{coupling}}$) represents a three-mode hybrid system consisting of two photon modes and one magnon mode. Its eigenvalues give the complex eigenfrequencies of the hybridized modes, where the real parts correspond to resonance frequencies and the imaginary parts represent the linewidths. When the magnon interacts with the photon modes, hybridization leads to the formation of new eigenmodes. In the regime of coherent interaction, the real parts of the eigenfrequencies of the coupling matrix exhibit level repulsion, resulting in a splitting into upper and lower branches near resonance.

Because the above relations are written in the time domain, the corresponding frequency-domain equations are obtained by Fourier transformation (see supplementary material S3.2). The relation between input $p_{\text{in}}$ and output fields $p_{\text{out}}$ in frequency domain can be obtained as
\begin{equation}
p_{\text{out}}(\omega) - p_{\text{in}}(\omega) = -2i \left[
\sqrt{\gamma_1}\, a(\omega) + \sqrt{\gamma_2}\, b(\omega) + \sqrt{\gamma_3}\, c(\omega)
\right]
\tag{12}
\end{equation}

Using this input--output relation, the transmission profile can be written in matrix form as (see supplementary material S3.3) \cite{Shrivastava2024PhotonPhoton,Rao2021PerfectAbsorption}
\begin{equation}
S_{21} = 1 + K_{(1\times3)}^{T} \, M_{(3\times3)}^{-1} \, K_{(3\times1)}
\tag{13}
\end{equation}

where,
\[
K = \sqrt{2}
\begin{pmatrix}
\sqrt{\gamma_1} \\
\sqrt{\gamma_2} \\
\sqrt{\gamma_3}
\end{pmatrix},
\quad
\]
\[
\begin{pmatrix}
\omega - \tilde{\omega}_1' & -g_{12} + i\sqrt{\gamma_1 \gamma_2} & -g_{31} + i\sqrt{\gamma_3 \gamma_1} \\
-g_{12} + i\sqrt{\gamma_1 \gamma_2} & \omega - \tilde{\omega}_2' & -g_{23} + i\sqrt{\gamma_2 \gamma_3} \\
-g_{31} + i\sqrt{\gamma_3 \gamma_1} & -g_{23} + i\sqrt{\gamma_2 \gamma_3} & \omega - \tilde{\omega}_3'
\end{pmatrix}.
\]
The transmission spectrum contains both direct and indirect coupling terms with the coupling strength of 
$-g_{12} + i\sqrt{\gamma_1 \gamma_2}$, 
$-g_{23} + i\sqrt{\gamma_2 \gamma_3}$, and 
$-g_{31} + i\sqrt{\gamma_3 \gamma_1}$, and interference between three channels with input field $p_{\text{in}}$ via $\sqrt{\gamma_1}$, $\sqrt{\gamma_2}$, and $\sqrt{\gamma_3}$ also exists.

\subsection{Estimation of Photon–Magnon Coupling Strengths}

To further elucidate the three-mode interaction, theoretical transmission colormap plots calculated using Eq.~(13) are shown in Fig.~6(b). The spectra are accurately reproduced by taking an intrinsic magnon damping $\beta_3 = 1.0~\text{MHz}$ and a very small extrinsic damping $\gamma_3 = 0.01~\text{MHz}$. The extrinsic damping of the magnon mode remains weak because the microstrip transmission line primarily excites the ELCR, while the YIG film placed on top of the resonator couples only weakly to the traveling microwave field. The calculated colormap closely matches the experimentally measured spectra across the entire magnetic-field range, confirming the validity of the proposed three-mode model. Moreover, the theoretical results clearly show that the coupling between the magnon and the two photon modes varies with the rotation angle of the ELCR. This agreement between theory and experiment demonstrates that the ELCR--YIG device operates as a polarization-controlled three-mode photon--magnon hybrid system \cite{Harder2021CavityMagnonics}.

Having established the theoretical framework, we next use the model to quantitatively extract the photon--magnon coupling strengths from the experimental spectra. For this purpose, the eigenvalues of the coupling matrix ($H_{\text{coupling}}$) given in Eq.~(11) were fitted to the experimental colormap, with the resulting eigenvalue branches shown as black curves in Fig.~6(a). From these fits, the magnon--photon coupling strengths associated with photon mode-1 are obtained as $g_{31} = 56.5~\text{MHz}$, $80~\text{MHz}$, $98~\text{MHz}$, and $0.0~\text{MHz}$ for rotation angles of $0^\circ$, $30^\circ$, $60^\circ$ and $90^\circ$, respectively. In contrast, the coupling strengths between the magnon and photon mode-2 are $g_{23} = 0.0~\text{MHz}$, $76~\text{MHz}$, $50~\text{MHz}$, and $30~\text{MHz}$ at the same angles (see supplementary material Table~2). The photon--photon coupling term was retained in the general circuit formulation for completeness, but the fitted circuit parameters yielded $M_{12} \approx 0$ for all measured angles; accordingly, the direct inter-photon coupling was neglected in the three-mode Hamiltonian by setting $g_{12} = 0$. These results reveal a pronounced polarization dependence: within the measured angular range, the coupling associated with photon mode-1 increases from $0^\circ$ to $60^\circ$ and vanishes at $90^\circ$, whereas the coupling associated with photon mode-2 emerges at finite angle and decreases from $30^\circ$ to $90^\circ$. This complementary behavior is consistent with the polarization-selective excitation of the resonator modes identified in the photon-only measurements discussed earlier.

The extracted coupling parameters further allow us to evaluate the cooperativity of the hybrid system, providing a quantitative measure of the photon--magnon interaction strength relative to dissipation. For the ELCR--YIG system, the cooperativities associated with photon mode-1 and photon mode-2 are defined as $C_1 = g_{31}^2/(K_1 \Gamma)$ and $C_2 = g_{23}^2/(K_2 \Gamma)$, where $K_1 = \beta_1 + \gamma_1$ and $K_2 = \beta_2 + \gamma_2$ represent the total linewidths of photon mode-1 and photon mode-2, respectively, and $\Gamma = \beta_3 + \gamma_3$ represents the total linewidth of the magnon mode \cite{Rameshti2022CavityMagnonics}. Here, all linewidth parameters are defined as half-width-at-half-maximum (HWHM) decay rates. Using the extracted coupling strengths and linewidth parameters, we obtain $C_1 \approx 226$, $480$, and $809$ for rotation angles $\theta = 0^\circ$, $30^\circ$, and $60^\circ$, respectively. Similarly, the cooperativity associated with photon mode-2 yields $C_2 \approx 201$, $71$, and $23$ for $\theta = 30^\circ$, $60^\circ$, and $90^\circ$, respectively. Since $C > 1$ throughout the measured angular range, coherent energy exchange between photons and magnons dominates over dissipative losses, confirming that the ELCR--YIG platform operates in the strong-coupling regime. The larger $C_1$ values indicate that, under the present conditions, photon mode-1 provides a more favorable coupling-to-loss balance than photon mode-2.

Building on the extracted coupling parameters, we further examine the relationship between the photon--magnon coupling strength and the extrinsic damping of the photon modes to better understand the underlying coupling mechanism. Figures~7(b) and 7(c) show the dependence of the coupling strength on the extrinsic damping for photon mode-1 and photon mode-2, respectively. For both modes, the coupling strength decreases as the extrinsic damping increases. However, this trend should not be interpreted as a direct suppression of the coupling by radiative damping. Instead, both quantities are governed by the same rotation-dependent field distribution. This complementary behavior reflects the polarization-selective excitation of the resonator modes and establishes a direct connection between the excitation polarization and the strength of photon--magnon hybridization.

\subsection{Polarization-Controlled Photon-Magnon Hybridization Channel Switching}

To obtain a comprehensive picture of polarization-controlled photon--magnon coupling in the ELCR-based hybrid system, we present the evolution of the coupling strengths $g_{31}$ (magenta diamonds) and $g_{23}$ (blue circles) together with the damping order parameter $\Phi_\gamma$ (red stars and black fitted curve) as a function of the rotation angle $\theta$ for our hybrid system as shown in Fig.~8. The two coupling channels exhibit complementary angular dependence, reflecting the orthogonality of the photon modes and resulting in a continuous redistribution of interaction strength as the resonator is rotated.

\begin{figure}
    \centering
    \includegraphics[width=\linewidth]{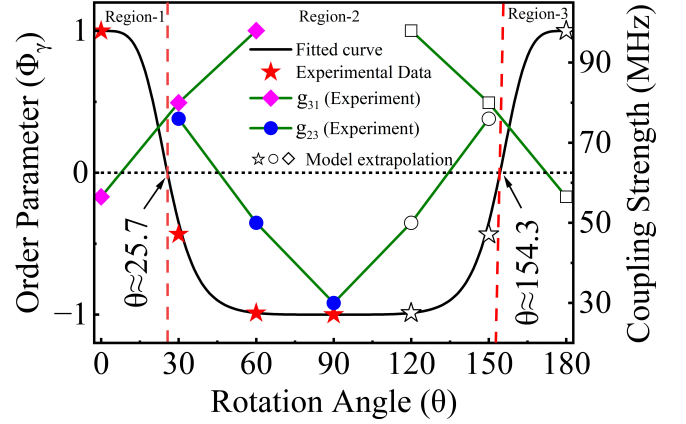}
    \caption{
Angular evolution of the radiative-loss imbalance and the two photon--magnon coupling channels. The damping order parameter $\Phi_\gamma$ is shown by red stars and the fitted angular dependence by the black curve. The extracted photon--magnon coupling strengths $g_{31}$ and $g_{23}$ are shown by magenta diamonds and blue circles, respectively. Within the measured range $0^\circ$--$90^\circ$, the data show a redistribution between the two hybridization channels as the resonator is rotated. The first critical angle, $\theta_{c1} \approx 25.7^\circ$, marks the measured crossing where $\Phi_\gamma = 0$. The extension to larger angles (shown as black hollow symbols), including the second crossing near $\theta_{c2} \approx 154.3^\circ$, represents a symmetry-related extrapolation of the fitted order-parameter model and is not directly established by the present measurements.
}
\label{F2}
\end{figure}

A key feature revealed in this figure is the distinct and nontrivial interplay between radiative damping and photon--magnon coupling. For rotation angles below the first critical angle, $\theta_{c1} \approx 25.7^\circ$ (region-1), the radiative damping is dominated by photon mode-1 ($\gamma_1 > \gamma_2$), as indicated by the positive values of the order parameter $\Phi_\gamma$. In this regime, photon mode-1 governs the hybridization, with $g_{31}$ finite while photon mode-2 is not excited and therefore does not participate in the coupling process. At $\theta_{c1}$, where $\Phi_\gamma = 0$, the fitted damping contributions become equal, marking the measured transition between the two hybridization channels within the experimentally accessible angular range. In the intermediate range $\theta_{c1} < \theta < \theta_{c2}$ (region-2), the roles of the two modes are reversed. Photon mode-2 becomes the dominant radiative channel ($\gamma_2 > \gamma_1$), and this increased radiative damping is accompanied by a redistribution of the microwave field that reduces its overlap with the magnon mode, resulting in $g_{23} < g_{31}$. This inversion indicates that the mode with stronger radiative damping corresponds to weaker photon--magnon coupling due to their common geometric dependence. By symmetry of the fitted order-parameter model, a second crossing appears near $\theta_{c2} \approx 154.3^\circ$; however, this lies outside the measured range and should therefore be interpreted as a symmetry-related, model-predicted transition rather than a directly observed one. For $\theta > \theta_{c2}$ (region-3), the system re-enters a regime similar to region-1, where photon mode-1 again dominates the radiative response, while the relative coupling strengths are determined by the same underlying geometric redistribution of the microwave field. These two critical angles, $\theta_{c1}$ and $\theta_{c2}$, define symmetric switching boundaries that partition the full angular range into distinct regimes characterized by polarization-controlled redistribution between the two competing photon--magnon hybridization channels.

The good agreement between the experimental data and the fitted curve indicates that this behavior arises from projection-controlled interaction between the excitation field and the orthogonal current-loop modes of the resonator. Rotation modifies the field overlap with each mode, thereby independently tuning both radiative losses and photon--magnon coupling strengths. Importantly, this analysis reveals that radiative damping and coupling strength are not inherently correlated but can be selectively controlled---and even inverted---through polarization rotation. Such controlled switching between hybridization channels provides a useful approach for engineering light--matter interactions in planar cavity--magnonic systems.

\section{Conclusion }

In this work, we have demonstrated polarization-controlled photon mode switching and photon–magnon coupling in a planar electric–LC resonator integrated with a YIG thin film. By rotating the resonator with respect to the fixed microwave field polarization of the transmission line, the excitation of two orthogonal photon modes can be selectively controlled, leading to a redistribution of the effective interaction between two competing channels. The equivalent circuit model successfully captures the polarization-dependent excitation and suppression of the two photon modes, providing a physically transparent description of the observed mode-switching behavior. In contrast, the effective three-mode Hamiltonian formalism is employed to describe the photon–magnon interaction, accurately reproducing the hybridized mode evolution observed in the transmission spectra. The extracted coupling strengths show that the two photon–magnon interaction channels are activated over different angular ranges and are redistributed with resonator orientation, reflecting polarization-controlled hybridization rather than a simple monotonic increase of a single coupling pathway. To quantify this behavior, a phenomenological order parameter is introduced to characterize the relative dominance of the two coupling channels. These results establish resonator orientation as an effective control parameter for both photon mode excitation and photon–magnon hybridization in planar systems. This approach provides a flexible route for controlling multimode interactions in cavity–magnonic platforms and may be useful for the development of reconfigurable magnonic and microwave devices.

\begin{acknowledgments}
This work was supported by Science and Engineering Research Board (SERB), India (Grant No.\ SRG/2023/001355), and the Anusandhan National Research Foundation (ANRF), India (Sanction Nos.\ ANRF/IRG/2025/001896/PS and ANRF/ARG/2025/006596/PS). Additional support was received from the Council of Science \& Technology, Uttar Pradesh (CSTUP) under Project IDs 2470 (Sanction No.\ CST/D-1520) and 4482 (Sanction No.\ CST/D-7/8). S.~Verma acknowledges the Ministry of Education, Government of India, for the Prime Minister’s Research Fellowship (PMRF ID-1102628). This research was further supported in part by the Basic Science Research Program through the National Research Foundation of Korea (NRF), funded by the Ministry of Science and ICT, South Korea (Grant No.\ RS-2024-00347921), with additional facilities provided by the Institute of Engineering Research at Seoul National University.
\end{acknowledgments}

\section*{Data Availability}
The data that support the findings of this study are available within the article.

\section*{supplementary}
\subsection*{S1. Scope of the supplementary theory}
This Supplementary Material provides the detailed derivations underlying the analytical descriptions used in the main text. In particular, it contains the derivation of the photon-only equivalent-circuit model, the definitions of intrinsic and extrinsic damping rates, and the input–output treatment of the three-mode photon–magnon hybrid system. The equations presented here support the compact main-text formulation used to analyze the polarization-controlled switching of the two photon modes and the corresponding redistribution of photon–magnon hybridization channels.

\subsection*{S2. Photon mode Dynamics}
\subsubsection*{S2.1 Photon mode-1}
Photon mode-1 consists of two dominant circulating current loops. Due to the spatially nonuniform microwave magnetic near field of the microstrip, the two loops experience unequal magnetic flux, resulting in inductive couplings of opposite sign (shown in Fig.~2(a) main text). Denoting the contributions from the two loops as $M_{1\theta}^{(1)}$ and $M_{1\theta}^{(2)}$, the effective mutual inductance of photon mode-1 is given by \cite{Naqui2014ModeSuppression}
\begin{equation}
M_1(\theta) = \frac{M_{1\theta}^{(1)} + M_{1\theta}^{(2)}}{2}
\tag{S1}
\end{equation}

At $\theta = 0^\circ$, the asymmetry in the field distribution is maximized, leading to $|M_{1\theta}^{(1)}| \neq |M_{1\theta}^{(2)}|$ and hence a finite net inductive coupling, i.e., $M_1(0^\circ) \neq 0$. At $\theta = 90^\circ$, the two loop contributions become equal in magnitude and opposite in sign,
\[
M_{1\theta}^{(1)} = -M_{1\theta}^{(2)},
\]
resulting in $M_1(90^\circ) = 0$. This explains why photon mode-1 is bright at $\theta = 0^\circ$ and dark at $\theta = 90^\circ$.
\subsubsection*{S2.2 Photon mode-2}
Photon mode-2 consists of four dominant circulating current loops, which can be grouped into two pairs with opposite effective circulation relative to the excitation field (shown in Fig.~2(b) main text). Denoting the four contributions as $M_{2\theta}^{(1)}$, $M_{2\theta}^{(2)}$, $M_{2\theta}^{(3)}$, and $M_{2\theta}^{(4)}$, the effective mutual inductance of photon mode-2 is given by
\begin{equation}
M_2(\theta) = \frac{(M_{2\theta}^{(1)} + M_{2\theta}^{(2)}) + (M_{2\theta}^{(3)} + M_{2\theta}^{(4)})}{2}
\tag{S2}
\end{equation}

At $\theta = 0^\circ$, the magnetic flux threading the loops is symmetric, leading to cancellation between the two pairs,
\[
(M_{2\theta}^{(1)} + M_{2\theta}^{(2)}) = -(M_{2\theta}^{(3)} + M_{2\theta}^{(4)}),
\]
and hence $M_2(0^\circ) = 0$.
At $\theta = 90^\circ$, this symmetry is broken, resulting in $M_2(90^\circ) \neq 0$. Therefore, photon mode-2 is dark at $\theta = 0^\circ$ and bright at $\theta = 90^\circ$. The complementary angular dependence of $M_1(\theta)$ and $M_2(\theta)$ gives rise to polarization-controlled mode switching.

\subsubsection*{S2.3 Equivalent circuit model of the photon-only system}

The photon-only response of the microstrip-loaded ELCR is modeled using a lumped-element equivalent circuit. The microstrip line is represented as a two-port network with characteristic impedance $Z_0$, while the two photon modes are modeled as two RLC resonators coupled to the line through mutual inductances $M_1$ and $M_2$ for photon mode-1 and mode-2, respectively. A possible direct inter-resonator mutual inductance $M_{12}$ is also included at the general level (as shown in Fig.~3(a) in main text). Let $V_L$, $V_{L1}$, and $V_{L2}$ denote the induced voltages in the transmission line, resonator-1, and resonator-2, respectively, and let $I$, $I_1$, and $I_2$ denote the corresponding currents.

The induced voltages can be written using Kirchhoff’s circuit laws as \cite{Pozar2021MicrowaveEngineering},
\begin{equation}
V_L = j\omega L I + j\omega M_1 I_1 + j\omega M_2 I_2
\tag{S3}
\end{equation}
\begin{equation}
V_{L1} = j\omega L_1 I_1 + j\omega M_1 I + j\omega M_{12} I_2
\tag{S4}
\end{equation}
\begin{equation}
V_{L2} = j\omega L_2 I_2 + j\omega M_2 I + j\omega M_{12} I_1
\tag{S5}
\end{equation}

For resonator-1, the sum of the voltages across the individual elements satisfies Kirchhoff's voltage law,
$V_{L1}+V_{R1}+V_{C1}=0$,
where \(V_{L1}\) is the induced voltage across the inductor \(L_1\), \(V_{R1}=I_1R_1\) is the voltage across the resistor \(R_1\), and
$V_{C1}=\frac{I_1}{j\omega C_1}$
is the voltage across the capacitor \(C_1\). So,
\[
j\omega L_1 I_1 + j\omega M_1 I + j\omega M_{12} I_2 + I_1 R_1 + \frac{I_1}{j\omega C_1} = 0
\]
or
\begin{equation}
\left(\frac{1}{C_1} + j\omega R_1 - \omega^2 L_1 \right) I_1 - \omega^2 M_{12} I_2 = \omega^2 M_1 I
\tag{S6}
\end{equation}

Similarly, for photon mode-2, the sum of the induced voltages across the individual elements is \(V_{L2}+V_{R2}+V_{C2}=0\), so
\begin{equation}
-\omega^2 M_{12} I_1 + \left(\frac{1}{C_2} + j\omega R_2 - \omega^2 L_2 \right) I_2 = \omega^2 M_2 I
\tag{S7}
\end{equation}

Since the resonance frequencies of the photon modes are given by \(\omega_1=\frac{1}{\sqrt{L_1C_1}}\) and \(\omega_2=\frac{1}{\sqrt{L_2C_2}}\) for photon mode-1 and mode-2, respectively, Eqs.~(S6) and (S7) can be simplified as
\begin{equation}
(L_1 \omega_1^2 - \omega^2 L_1 + j\omega R_1) I_1 - \omega^2 M_{12} I_2 = \omega^2 M_1 I
\tag{S8}
\end{equation}
\begin{equation}
-\omega^2 M_{12} I_1 + (L_2 \omega_2^2 - \omega^2 L_2 + j\omega R_2) I_2 = \omega^2 M_2 I
\tag{S9}
\end{equation}

Solving the linear system formed by Eqs.~(S8) and (S9) for $I_1$ and $I_2$, and substituting the result into Eq.~(S3), yields the induced voltage in the transmission line \cite{Aznar2008LeftHandedLines},
\begin{widetext}
\begin{equation}
V_L = j\omega L I + \frac{
j\omega^3 \left\{
C_1 M_1^2 \left(1 - \frac{\omega^2}{\omega_2^2}\right)
+ C_2 M_2^2 \left(1 - \frac{\omega^2}{\omega_1^2}\right)
+ 2\omega M_{12} M_1 M_2 C_1 C_2
+ j\omega C_1 C_2 (R_1 M_2^2 + R_2 M_1^2)
\right\}
}{
\left(1 - \frac{\omega^2}{\omega_1^2} + j\omega R_1 C_1 \right)
\left(1 - \frac{\omega^2}{\omega_2^2} + j\omega R_2 C_2 \right)
- \omega^4 M_{12}^2 C_1 C_2
} I
\tag{S10}
\end{equation}
\end{widetext}
The total series impedance seen by the transmission line is therefore $Z_s(\omega) = V_L/I$, which, using Eq.~(S10), can be written as,
\begin{equation}
Z_s(\omega) = j\omega L + \Delta Z(\omega)
\tag{S11}
\end{equation}

Here,
\begin{widetext}
\[
\Delta Z(\omega) =
\frac{
j\omega^3 \left\{
C_1 M_1^2 \left(1 - \frac{\omega^2}{\omega_2^2}\right)
+ C_2 M_2^2 \left(1 - \frac{\omega^2}{\omega_1^2}\right)
+ 2\omega M_{12} M_1 M_2 C_1 C_2
+ j\omega C_1 C_2 (R_1 M_2^2 + R_2 M_1^2)
\right\}
}{
\left(1 - \frac{\omega^2}{\omega_1^2} + j\omega R_1 C_1 \right)
\left(1 - \frac{\omega^2}{\omega_2^2} + j\omega R_2 C_2 \right)
- \omega^4 M_{12}^2 C_1 C_2
}.
\]
\end{widetext}

This expression gives the equivalent series impedance when the resonators are coupled to the transmission line. To compute the transmission coefficient $S_{21}$, we employ the standard ABCD-matrix formalism for cascaded two-port networks. The equivalent circuit can be decomposed into three elements: a central series impedance $Z_s$, and two identical shunt admittances $Y_1$ and $Y_2$, corresponding to the capacitive loading of the transmission line (as shown in Fig.~3(b) in main text). These admittances are defined as $Y_1 = i\omega C/2$ and $Y_2 = i\omega C/2$. The overall ABCD matrix of the system is obtained by multiplying the matrices of the individual elements in cascade. From the resulting ABCD parameters, the transmission coefficient $S_{21}$ is calculated using standard relations for two-port networks \cite{Pozar2021MicrowaveEngineering,Kaur2016EITMagnonics}.

\begin{equation}
\begin{pmatrix}
A & B \\
C & D
\end{pmatrix}
=
\begin{pmatrix}
1 & 0 \\
Y_1 & 1
\end{pmatrix}
\begin{pmatrix}
1 & Z_s \\
0 & 1
\end{pmatrix}
\begin{pmatrix}
1 & 0 \\
Y_2 & 1
\end{pmatrix}
\tag{S12}
\end{equation}

By simplifying the above matrix, we obtain the ABCD parameters as,
\[
\begin{aligned}
A &= 1 + Y_2 Z_s, \quad
B = Z_s,\\
C &= Y_1 + Y_2 (1 + Y_1 Z_s), \quad
D = 1 + Y_1 Z_s
\end{aligned}
\]

The $S_{21}$ parameter of the two-port network can then be written in terms of the ABCD parameters as follows \cite{Pozar2021MicrowaveEngineering,Aznar2008LeftHandedLines}:
\begin{equation}
S_{21} = \frac{2}{A + \frac{B}{Z_0} + C Z_0 + D}
\tag{S14}
\end{equation}

The above expression for $S_{21}$ applies to the two-resonator system, where $Z_0$ is the characteristic impedance of the transmission line; in the present case, $Z_0 = 50~\Omega$.

\subsubsection*{S2.4 Intrinsic and extrinsic damping parameters}
Damping originates from energy dissipation in the system. In the equivalent circuit description, this dissipation is captured by the real part of the impedance $\mathrm{Re}[Z_s(\omega)]$, which quantifies power loss. This loss translates into a finite linewidth (damping rate) in the frequency response.

\textbf{Intrinsic Damping}

When the resonators are decoupled from the transmission line, i.e., $M_1 = M_2 = M_{12} = 0$, the resonator dynamics are governed by the denominator,
\begin{equation}
D = \left(1 - \frac{\omega^2}{\omega_1^2} + j\omega R_1 C_1 \right)
\left(1 - \frac{\omega^2}{\omega_2^2} + j\omega R_2 C_2 \right)
\tag{S15}
\end{equation}

Near photon mode-1 resonance,
\begin{equation}
\left(1 - \frac{\omega^2}{\omega_1^2} \right) \approx -\frac{2}{\omega_1} (\omega - \omega_1)
\tag{S16}
\end{equation}

So, the denominator becomes for the photon mode-1 system,
\begin{equation}
\left(1 - \frac{\omega^2}{\omega_1^2} + j\omega R_1 C_1 \right)
\approx -\frac{2}{\omega_1} (\omega - \omega_1) + j\omega_1 R_1 C_1
\tag{S17}
\end{equation}

To extract a linewidth, compare with the standard Lorentzian, $(\omega - \omega_1) + j\beta_1$, where $\beta_1$ is the intrinsic damping of photon mode-1, so
\begin{equation}
\beta_1 = \frac{\omega_1^2 R_1 C_1}{2}
\tag{S18}
\end{equation}

Similarly for photon mode-2 intrinsic damping,
\begin{equation}
\beta_2 = \frac{\omega_2^2 R_2 C_2}{2}
\tag{S19}
\end{equation}

\textbf{Extrinsic Damping}

Since the power radiated into the transmission line is 
$P = \frac{|V_{\mathrm{ind}}|^2}{2Z_0}$
and the energy stored in the LC resonator is 
$U = \frac{1}{2} L I^2,$
the radiative damping is given by,
\[
\gamma = \frac{P}{2U}
\]

\begin{equation}
\gamma = \frac{|V_{\mathrm{ind}}|^2}{2 Z_0 L I^2}
\tag{S20}
\end{equation}

Since, for photon mode-1 the induced voltage is $V_{\mathrm{ind}} = j\omega M_1 I_1$ and the resonance frequency $\omega_1 = \frac{1}{\sqrt{L_1 C_1}}$, the extrinsic damping for photon mode-1 is given as,
\begin{equation}
\gamma_1 = \frac{\omega_1^4 M_1^2 C_1}{2Z_0}
\tag{S21}
\end{equation}

And similarly, for photon mode-2,
\begin{equation}
\gamma_2 = \frac{\omega_2^4 M_2^2 C_2}{2Z_0}
\tag{S22}
\end{equation}\\

\subsection*{S3. Photon–Magnon Dynamics}

\subsubsection*{S3.1 Three-mode Hamiltonian for the ELCR–YIG platform}
The ELCR--YIG system is modeled as two photon modes coupled to one magnon mode and to a common waveguide continuum as shown in Fig.~7(a). Before the rotating-wave approximation, the Hamiltonian is written as \cite{Harder2021CavityMagnonics}
\begin{widetext}
\begin{align}
\frac{H}{\hbar} &= \tilde{\omega}_1 a^\dagger a + \tilde{\omega}_2 b^\dagger b + \tilde{\omega}_3 c^\dagger c \nonumber + g_{12} (a + a^\dagger)(b + b^\dagger) + g_{23} (b + b^\dagger)(c + c^\dagger)
+ g_{31} (c + c^\dagger)(a + a^\dagger) \nonumber
+ \\&\int_{-\infty}^{+\infty} \omega_k p_k^\dagger p_k \, dk \nonumber + \int_{-\infty}^{+\infty}
\Big[
\lambda_1 \cos\theta (a + a^\dagger)(p_k + p_k^\dagger)
+ \lambda_2 \sin\theta (b + b^\dagger)(p_k + p_k^\dagger)
+ \lambda_3 (c + c^\dagger)(p_k + p_k^\dagger)
\Big] dk
\tag{23}
\end{align}
\end{widetext}

The operators $a$, $b$, and $c$ denote the annihilation operators of photon mode-1, photon mode-2, and the magnon mode, respectively, while $p_k$ represents the continuum of traveling photon modes in the waveguide. The resonance frequencies of the uncoupled modes are represented by the complex quantities $\tilde{\omega}_1$, $\tilde{\omega}_2$, and $\tilde{\omega}_3$, defined as $\tilde{\omega}_1 = \omega_1 - i\beta_1$, $\tilde{\omega}_2 = \omega_2 - i\beta_2$, and $\tilde{\omega}_3 = \omega_3 - i\beta_3$. Here $\beta_1$, $\beta_2$, and $\beta_3$ represent the intrinsic damping rates, and $\lambda_1$, $\lambda_2$, and $\lambda_3$ denote the coupling strengths to the waveguide continuum, for the photon mode-1, photon mode-2, and the magnon mode, respectively. The angle dependence $\cos\theta$ and $\sin\theta$ arises from the polarization overlap between the excitation field and the photon modes.

Since the above Hamiltonian contains non-conserving terms, we apply the rotating-wave approximation (RWA), under which the Hamiltonian reduces to \cite{Zhang2016Magnomechanics}
\begin{widetext}
\begin{align}
\frac{H}{\hbar} &= \tilde{\omega}_1 a^\dagger a + \tilde{\omega}_2 b^\dagger b + \tilde{\omega}_3 c^\dagger c \nonumber + g_{12} (a^\dagger b + a b^\dagger)
+ g_{23} (b^\dagger c + b c^\dagger)
+ g_{31} (c^\dagger a + c a^\dagger) \nonumber + \int_{-\infty}^{+\infty} \omega_k p_k^\dagger p_k \, dk \nonumber \\&+ \int_{-\infty}^{+\infty}
\Big[
\lambda_1 \cos\theta (a^\dagger p_k + a p_k^\dagger)
+ \lambda_2 \sin\theta (b^\dagger p_k + b p_k^\dagger)
+ \lambda_3 (c^\dagger p_k + c p_k^\dagger)
\Big] dk
\tag{24}
\end{align}
\end{widetext}

\subsubsection*{S3.2 Heisenberg–Langevin Formalism and Mode Dynamics}
Now using the Heisenberg--Langevin, the equation of motion of the traveling photon is given as,

\begin{align}
\dot{p}_k(t) &= \frac{-i}{\hbar}[p_k, H] = -i\omega_k p_k - i\lambda_1 (\cos\theta)a - i\lambda_2 (\sin\theta)b \nonumber \\
&\quad - i\lambda_3 c
\tag{S25}
\end{align}

Importantly, the angular structure in Eq.~(S24) is preserved throughout the derivation. In particular, the mode-1 channel always carries the factor $\cos\theta$, the mode-2 channel always carries the factor $\sin\theta$, and the magnon channel carries no rotation factor. Equation~(S25) can be solved by considering the forward-time solution for the traveling photon,

\begin{align}
p_k(t) &= e^{-i\omega_k (t - t_0)} p_k(t_0) 
- \int_{t_0}^{t} i\left[\lambda_1 (\cos\theta)a(t')\right. \nonumber \\
&\quad \left. + \lambda_2 (\sin\theta)b(t')  + \lambda_3 c(t')\right] e^{-i\omega_k (t - t')} dt'
\tag{S26}
\end{align}

Where $p_k(t_0)$ is the initial state of $p_k(t)$ ($t_0 < t$). The input field operator at input port is defined as,
\begin{align}
p_{\text{in}}(t) = \frac{1}{\sqrt{2\pi}} \int e^{-i\omega_k (t - t_0)} p_k(t_0)\, dk
\tag{S27}
\end{align}

The equation of motion of photon mode-1 is given by,
\begin{align}
\dot{a}(t) &= \frac{-i}{\hbar}[a,H] = -i\tilde{\omega}_1 a(t) - ig_{12} b(t) - ig_{31} c(t) \nonumber \\
&\quad - i \int_{-\infty}^{\infty} \lambda_1 \cos\theta\, p_k(t)\, dk
\tag{S28}
\end{align}

Within the first Markov approximation, $\lambda_{1,2,3}$ may be treated as constants and taken outside the $k$-integral. Substituting Eqs.~(S26) and (S27) into Eq.~(S28) yields,
\begin{align}
\dot{a}(t) &= -i\tilde{\omega}_1 a(t) - ig_{12} b(t) - ig_{31} c(t) - i\sqrt{\gamma_1} p_{\text{in}}(t)
- \gamma_1 a(t) \nonumber \\
&\quad - \sqrt{\gamma_1 \gamma_2} b(t) - \sqrt{\gamma_3 \gamma_1} c(t)
\tag{S29}
\end{align}

Similarly, for the mode $b$ and $c$,
\begin{align}
\dot{b}(t) &= -i\tilde{\omega}_2 b(t) - ig_{12} a(t) - ig_{23} c(t) - i\sqrt{\gamma_2} p_{\text{in}}(t)\nonumber \\
&\quad - \sqrt{\gamma_1 \gamma_2} a(t) - \gamma_2 b(t) - \sqrt{\gamma_2 \gamma_3} c(t)
\tag{S30}
\end{align}
\begin{align}
\dot{c}(t) &= -i\tilde{\omega}_3 c(t) - ig_{23} b(t) - ig_{31} a(t) - i\sqrt{\gamma_3} p_{\text{in}}(t)\nonumber \\
&\quad - \sqrt{\gamma_3 \gamma_1} a(t) - \sqrt{\gamma_2 \gamma_3} b(t) - \gamma_3 c(t)
\tag{S31}
\end{align}

Whereas $\gamma_1 = 2\pi \lambda_1^2 \cos^2\theta$, $\gamma_2 = 2\pi \lambda_2^2 \sin^2\theta$, and $\gamma_3 = 2\pi \lambda_3^2$ are the extrinsic damping of the modes photon mode-1, photon mode-2, and magnon mode, respectively. Here $\gamma_1(0) = 2\pi \lambda_1^2$ and $\gamma_2(90^\circ) = 2\pi \lambda_2^2$, so extrinsic damping can be written in the form of rotation angle ($\theta$) dependence as $\gamma_1 = \gamma_1(0)\cos^2\theta$ and $\gamma_2 = \gamma_2(90^\circ)\sin^2\theta$ for the photon mode-1 and photon mode-2, respectively.

The Fourier transforms of Eqs.~(S29)--(S31) are given by,
\begin{align}
&i(\omega - \tilde{\omega}_1)a(\omega) - ig_{12} b(\omega) - ig_{31} c(\omega) - i\sqrt{\gamma_1} p_{\text{in}}(\omega)\nonumber \\
&\quad - \gamma_1 a(\omega)  - \sqrt{\gamma_1 \gamma_2} b(\omega) - \sqrt{\gamma_3 \gamma_1} c(\omega) = 0
\tag{S32}
\end{align}
\begin{align}
&i(\omega - \tilde{\omega}_2)b(\omega) - ig_{12} a(\omega) - ig_{23} c(\omega) - i\sqrt{\gamma_2} p_{\text{in}}(\omega)
\nonumber \\
&\quad - \sqrt{\gamma_1 \gamma_2} a(\omega) - \gamma_2 b(\omega) - \sqrt{\gamma_2 \gamma_3} c(\omega) = 0
\tag{S33}
\end{align}
\begin{align}
&i(\omega - \tilde{\omega}_3)c(\omega) - ig_{23} b(\omega) - ig_{31} a(\omega) - i\sqrt{\gamma_3} p_{\text{in}}(\omega)
\nonumber \\
&\quad - \sqrt{\gamma_3 \gamma_1} a(\omega) - \sqrt{\gamma_2 \gamma_3} b(\omega) - \gamma_3 c(\omega) = 0
\tag{S34}
\end{align}

We can also solve Eq.~(S24) to write $p_k(t)$ in the terms of time-retarded equation for the time $t_1 > t$ to calculate transmission spectra,
\begin{align}
&p_k(t) = e^{-i\omega_k (t - t_1)} p_k(t_1) + \int_{t}^{t_1} i\left[\lambda_1 (\cos\theta)a(t') \right. \nonumber \\
&\quad \left. +  \lambda_2 (\sin\theta)b(t') + \lambda_3 c(t')\right] e^{-i\omega_k (t - t')} dt'
\tag{S35}
\end{align}

Where $p_k(t_1)$ is the late time state of $p_k(t)$ as $t_1 > t$. The output field operator at output port is defined as,
\begin{equation}
p_{\text{out}}(t) = \frac{1}{\sqrt{2\pi}} \int e^{-i\omega_k (t - t_1)} p_k(t_1)\, dk
\tag{S36}
\end{equation}

Using Eqs.~(S35) and (S36) in Eq.~(S28), the coupled Heisenberg--Langevin equations for photon mode-1, photon mode-2, and the magnon mode become,
\begin{align}
\dot{a}(t) &= -i\tilde{\omega}_1 a(t) - ig_{12} b(t) - ig_{31} c(t) - i\sqrt{\gamma_1} p_{\text{out}}(t) \nonumber \\
&\quad + \gamma_1 a(t) +  \sqrt{\gamma_1 \gamma_2} b(t) + \sqrt{\gamma_3 \gamma_1} c(t)
\tag{S37}
\end{align}
\begin{align}
\dot{b}(t) &= -i\tilde{\omega}_2 b(t) - ig_{12} a(t) - ig_{23} c(t) - i\sqrt{\gamma_2} p_{\text{out}}(t)
\nonumber \\
&\quad + \sqrt{\gamma_1 \gamma_2} a(t) + \gamma_2 b(t) + \sqrt{\gamma_2 \gamma_3} c(t)
\tag{S38}
\end{align}
\begin{align}
\dot{c}(t) &= -i\tilde{\omega}_3 c(t) - ig_{23} b(t) - ig_{31} a(t) - i\sqrt{\gamma_3} p_{\text{out}}(t)
\nonumber \\
&\quad + \sqrt{\gamma_3 \gamma_1} a(t) + \sqrt{\gamma_2 \gamma_3} b(t) + \gamma_3 c(t)
\tag{S39}
\end{align}

The Fourier transforms of Eqs.~(S37)--(S39) from the time domain to the frequency domain are given by,
\begin{align}
&i(\omega - \tilde{\omega}_1)a(\omega) - ig_{12} b(\omega) - ig_{31} c(\omega) - i\sqrt{\gamma_1} p_{\text{out}}(\omega)\nonumber \\
&\quad
+ \gamma_1 a(\omega) + \sqrt{\gamma_1 \gamma_2} b(\omega) + \sqrt{\gamma_3 \gamma_1} c(\omega) = 0
\tag{S40}
\end{align}
\begin{align}
&i(\omega - \tilde{\omega}_2)b(\omega) - ig_{12} a(\omega) - ig_{23} c(\omega) - i\sqrt{\gamma_2} p_{\text{out}}(\omega)\nonumber \\
&\quad
+ \sqrt{\gamma_1 \gamma_2} a(\omega) + \gamma_2 b(\omega) + \sqrt{\gamma_2 \gamma_3} c(\omega) = 0
\tag{S41}
\end{align}
\begin{align}
&i(\omega - \tilde{\omega}_3)c(\omega) - ig_{23} b(\omega) - ig_{31} a(\omega) - i\sqrt{\gamma_3} p_{\text{out}}(\omega)\nonumber \\
&\quad
+ \sqrt{\gamma_3 \gamma_1} a(\omega) + \sqrt{\gamma_2 \gamma_3} b(\omega) + \gamma_3 c(\omega) = 0
\tag{S42}
\end{align}

\subsubsection*{S3.3 Transmission Spectrum of the Hybrid System}

Using Eqs.~(S29)--(S31), the Heisenberg--Langevin equations can be written compactly in matrix form as \cite{Shrivastava2024PhotonPhoton}:
\begin{equation}
\frac{d}{dt}
\begin{pmatrix}
a(t) \\ b(t) \\ c(t)
\end{pmatrix}
=
-i H_{\text{coupling}}
\begin{pmatrix}
a(t) \\ b(t) \\ c(t)
\end{pmatrix}
-i
\begin{pmatrix}
\sqrt{\gamma_1} \\ \sqrt{\gamma_2} \\ \sqrt{\gamma_3}
\end{pmatrix}
p_{\text{in}}(t)
\tag{S43}
\end{equation}

Where the coupling matrix is given by,
\begin{equation}
H_{\text{coupling}} =
\begin{pmatrix}
\tilde{\omega}_1' & g_{12} - i\sqrt{\gamma_1 \gamma_2} & g_{31} - i\sqrt{\gamma_3 \gamma_1} \\
g_{12} - i\sqrt{\gamma_1 \gamma_2} & \tilde{\omega}_2' & g_{23} - i\sqrt{\gamma_2 \gamma_3} \\
g_{31} - i\sqrt{\gamma_3 \gamma_1} & g_{23} - i\sqrt{\gamma_2 \gamma_3} & \tilde{\omega}_3'
\end{pmatrix}
\tag{S44}
\end{equation}

The relation between $p_{\text{in}}(\omega)$ and $p_{\text{out}}(\omega)$ follows from the frequency-domain mode equations together with the input--output relation, obtained by combining Eqs.~(S32) and (S40), as,
\begin{equation}
p_{\text{out}}(\omega) - p_{\text{in}}(\omega) = -2i \left[
\sqrt{\gamma_1} a(\omega) + \sqrt{\gamma_2} b(\omega) + \sqrt{\gamma_3} c(\omega)
\right]
\tag{S45}
\end{equation}

Because the above equations are cumbersome to solve analytically, it is convenient to use a matrix formulation for the transmission coefficient $S_{21}$. The input--output relation in Eq.~(S45) can then be rewritten as,
\begin{equation}
p_{\text{out}}(\omega) - p_{\text{in}}(\omega) =
-2i
\begin{pmatrix}
\sqrt{\gamma_1} & \sqrt{\gamma_2} & \sqrt{\gamma_3}
\end{pmatrix}
\begin{pmatrix}
a(\omega) \\ b(\omega) \\ c(\omega)
\end{pmatrix}
\tag{S46}
\end{equation}

The transmission coefficient $S_{21}$ between input and output port can be written as \cite{Shrivastava2024PhotonPhoton,Rameshti2022CavityMagnonics,Rao2020TravelingPhotons},
\begin{equation}
S_{21} = \frac{p_{\text{out}}(\omega)}{p_{\text{in}}(\omega)}
\tag{S47}
\end{equation}

Now Eqs.~(S32), (S33), and (S34) can be formulated in matrix form as follows,
\begin{align}
&\begin{pmatrix}
i(\omega - \tilde{\omega}_1) - \gamma_1 & -ig_{12} - \sqrt{\gamma_1 \gamma_2} & -ig_{31} - \sqrt{\gamma_3 \gamma_1} \\
-ig_{12} - \sqrt{\gamma_1 \gamma_2} & i(\omega - \tilde{\omega}_2) - \gamma_2 & -ig_{23} - \sqrt{\gamma_2 \gamma_3} \\
-ig_{31} - \sqrt{\gamma_3 \gamma_1} & -ig_{23} - \sqrt{\gamma_2 \gamma_3} & i(\omega - \tilde{\omega}_3) - \gamma_3
\end{pmatrix}
\begin{pmatrix}
a(\omega) \\ b(\omega) \\ c(\omega)
\end{pmatrix}
\nonumber \\
&\quad - i
\begin{pmatrix}
\sqrt{\gamma_1} \\ \sqrt{\gamma_2} \\ \sqrt{\gamma_3}
\end{pmatrix}
p_{\text{in}}(\omega)
= 0
\tag{S48}
\end{align}

Using Eqs.~(S46)--(S48), the transmission profile can be written in matrix form as,
\begin{equation}
S_{21} = 1 + K_{(1\times3)}^{T} M_{(3\times3)}^{-1} K_{(3\times1)}
\tag{S49}
\end{equation}

Whereas,
\begin{equation}
K = \sqrt{2}
\begin{pmatrix}
\sqrt{\gamma_1} \\ \sqrt{\gamma_2} \\ \sqrt{\gamma_3}
\end{pmatrix}
\tag{S50}
\end{equation}

\begin{equation}
M = i
\begin{pmatrix}
\omega - \tilde{\omega}_1' & -g_{12} + i\sqrt{\gamma_1 \gamma_2} & -g_{31} + i\sqrt{\gamma_3 \gamma_1} \\
-g_{12} + i\sqrt{\gamma_1 \gamma_2} & \omega - \tilde{\omega}_2' & -g_{23} + i\sqrt{\gamma_2 \gamma_3} \\
-g_{31} + i\sqrt{\gamma_3 \gamma_1} & -g_{23} + i\sqrt{\gamma_2 \gamma_3} & \omega - \tilde{\omega}_3'
\end{pmatrix}
\tag{S51}
\end{equation}

Where $\tilde{\omega}_1'$, $\tilde{\omega}_2'$, and $\tilde{\omega}_3'$ are the resultant complex frequencies of the three modes respectively, i.e. $\tilde{\omega}_1' = \omega_1 - i(\beta_1 + \gamma_1)$, $\tilde{\omega}_2' = \omega_2 - i(\beta_2 + \gamma_2)$, and $\tilde{\omega}_3' = \omega_3 - i(\beta_3 + \gamma_3)$. The transmission spectrum contains both direct and indirect coupling terms with the coupling strength of $-g_{12} + i\sqrt{\gamma_1 \gamma_2}$, $-g_{23} + i\sqrt{\gamma_2 \gamma_3}$ and $-g_{31} + i\sqrt{\gamma_3 \gamma_1}$ and interference between three channels with input field $p_{\text{in}}$ via $\sqrt{\gamma_1}$, $\sqrt{\gamma_2}$, and $\sqrt{\gamma_3}$ also exists.

\subsection*{S4. Fitting strategy and parameter definitions}
The fitting procedure was carried out in two stages.

In the first stage, the photon-only transmission spectra were fitted using the equivalent circuit model [Eq.~(S14)]. This allowed us to extract the resonance frequencies ($\omega_1$, $\omega_2$), capacitances ($C_1$, $C_2$), mutual inductances ($M_1$, $M_2$), resistances ($R_1$, $R_2$) of the two photon modes, as well as the inter-mode mutual inductance ($M_{12}$) and the transmission line parameters (capacitance $C$ and inductance $L$). The extracted parameters are summarized in Table~1. Using these values, the intrinsic damping was calculated from Eqs.~(S18) and (S19), while the extrinsic damping was determined from Eqs.~(S21) and (S22). Here, both intrinsic and extrinsic damping are defined in terms of the half-width at half-maximum (HWHM) of the resonance peaks. The total linewidths of the modes are therefore given by $K_1 = \beta_1 + \gamma_1$ and $K_2 = \beta_2 + \gamma_2$ for the photon modes, and $\Gamma = \beta_3 + \gamma_3$ for the magnon mode. 

In the second stage, the hybrid photon--magnon spectra shown in Fig.~6(a) were fitted using the eigenvalues of Eq.~(S44). From this analysis, we extracted the coupling strengths between each photon mode and the magnon mode. In these fits, the photon-mode resonance frequencies, intrinsic damping, and extrinsic damping were fixed to the values obtained from the circuit model. All parameters used in fitting the eigenvalues of Eq.~(S44), corresponding to Fig.~6(a), as well as in generating the transmission colormap shown in Fig.~6(b) plotted using Eq.~(S49), are listed in Table~2.

The direct photon--photon coupling was neglected ($g_{12} = 0$), as the photon-only circuit fitting yielded $M_{12} \approx 0$ and the spatial overlap between the two photon-mode field distributions was negligible. The extracted photon--magnon coupling strengths ($g_{31}$ and $g_{23}$) should therefore be understood as effective parameters within the hybrid model. Their angular dependence reflects a redistribution of interaction strength between two competing hybridization channels, rather than a strictly monotonic enhancement of one channel accompanied by suppression of the other.

\begin{table}[t]
\centering
\caption{Circuit parameters obtained by fitting the experimental two-mode $S_{21}$ transmission spectra using the equivalent circuit model.}
\begin{tabular}{lcccc}
\hline
Parameters & $0^\circ$ & $30^\circ$ & $60^\circ$ & $90^\circ$ \\
\hline
$\omega_1$ (GHz) & 3.9350 & 3.7557 & 3.8816 & 3.8816 \\
$\omega_2$ (GHz) & 5.6778 & 5.6778 & 5.7342 & 5.7138 \\
$C$ (pF)         & 1.2884 & 1.2884 & 1.2884 & 1.2884 \\
$C_1$ (pF)       & 0.2193 & 0.2193 & 0.2193 & 0.2193 \\
$C_2$ (pF)       & 0.2988 & 0.2988 & 0.2988 & 0.2988 \\
$L$ (nH)         & 0.9196 & 0.9196 & 0.9196 & 0.9196 \\
$M_1$ (nH)       & 0.2150 & 0.1840 & 0.1100 & 0.0000 \\
$M_2$ (nH)       & 0.0000 & 0.0930 & 0.1620 & 0.1820 \\
$M_{12}$ (nH)    & 0.0000 & 0.0000 & 0.0000 & 0.0000 \\
$R_1$ ($\Omega$) & 0.9831 & 0.9831 & 0.9831 & 0.9831 \\
$R_2$ ($\Omega$) & 0.8007 & 0.8007 & 0.8007 & 0.8007 \\
\hline
\end{tabular}
\end{table}

\begin{table}[t]
\centering
\caption{Parameters obtained by fitting the experimental $S_{21}$ transmission spectra using the quantum three-mode model.}
\begin{tabular}{lcccc}
\hline
Parameters & $0^\circ$ & $30^\circ$ & $60^\circ$ & $90^\circ$ \\
\hline
$\omega_1$ (GHz) & 3.9350 & 3.7557 & 3.8816 & 3.8816 \\
$\omega_2$ (GHz) & 5.6778 & 5.6778 & 5.7342 & 5.7138 \\
$\beta_3$ (MHz)  & 1.0    & 1.0    & 1.0    & 1.0 \\
$\beta_1$ (MHz)  & 11.0   & 11.0   & 11.0   & 11.0 \\
$\beta_2$ (MHz)  & 25.0   & 25.0   & 25.0   & 25.0 \\
$g_{12}$ (MHz)   & 0.0    & 0.0    & 0.0    & 0.0 \\
$g_{31}$ (MHz)   & 56.5   & 80.0   & 98.0   & 0.0 \\
$g_{23}$ (MHz)   & 0.0    & 76.0   & 50.0   & 30.0 \\
$\gamma_3$ (MHz) & 0.01   & 0.01   & 0.01   & 0.01 \\
$\gamma_1$ (MHz) & 3.0    & 2.2    & 0.75   & 0.0 \\
$\gamma_2$ (MHz) & 0.0    & 3.5    & 10.0   & 13.0 \\
\hline
\end{tabular}
\end{table}

\FloatBarrier

\nocite{*}
\bibliography{apssamp}


\bibliographystyle{apsrev4-2}
\end{document}